\documentclass{emulateapj}
\usepackage{color}

\begin{document}

\shortauthors{Luhman \& Esplin}
\shorttitle{Refining the Census of Upper Sco with {\it Gaia}}

\title{Refining the Census of the Upper Scorpius Association with {\it Gaia}\altaffilmark{1}}

\author{
K. L. Luhman\altaffilmark{2,3}
and T. L. Esplin\altaffilmark{4,5}
}

\altaffiltext{1}
{Based on observations made with the Two Micron All Sky Survey,
the {\it Gaia} mission, the NASA Infrared Telescope Facility,
Cerro Tololo Inter-American Observatory, Gemini Observatory, the
Magellan Baade Telescope at Las Campanas Observatory, and
the Bok Reflector at Steward Observatory.}

\altaffiltext{2}{Department of Astronomy and Astrophysics,
The Pennsylvania State University, University Park, PA 16802, USA;
kll207@psu.edu}
\altaffiltext{3}{Center for Exoplanets and Habitable Worlds, The
Pennsylvania State University, University Park, PA 16802, USA}
\altaffiltext{4}{Steward Observatory, University of Arizona, 933 North Cherry 
Avenue, Tucson, AZ 85721, USA}
\altaffiltext{5}{Strittmatter Fellow.}

\begin{abstract}

We have refined the census of stars and brown dwarfs in the Upper Sco
association ($\sim10$~Myr, $\sim145$~pc) by 1) updating the
selection of candidate members from our previous survey to include
the high-precision astrometry from the second data release of {\it Gaia},
2) obtaining spectra of a few hundred candidate members to measure
their spectral types and verify their youth, and 3) assessing
the membership (largely with {\it Gaia} astrometry) of 2020 stars toward
Upper Sco that show evidence of youth in this work and previous studies.
We arrive at a catalog of 1761 objects that are adopted as members of
Upper Sco.
The distribution of spectral types among the adopted members
is similar to those in other nearby star-forming regions,
indicating a similar initial mass function.
In previous studies, we have compiled mid-infrared photometry from
{\it WISE} and the {\it Spitzer Space Telescope} for members
of Upper Sco and used those data to identify the stars that show evidence
of circumstellar disks; we present the same analysis for our new catalog of
members. As in earlier work, we find that the
fraction of members with disks increases with lower stellar masses,
ranging from $\lesssim10$\% for $>1$~$M_\odot$ to $\sim22$\% for
0.01--0.3~$M_\odot$. Finally, we have estimated the relative ages of Upper Sco
and other young associations using their sequences of low-mass stars in
$M_{G_{\rm RP}}$ versus $G_{\rm BP}-G_{\rm RP}$.
This comparison indicates that Upper Sco is a factor of two younger than
the $\beta$~Pic association (21--24~Myr) according to both non-magnetic and
magnetic evolutionary models.

\end{abstract}


\section{Introduction}
\label{sec:intro}

The Upper Scorpius subgroup in the Scorpius-Centaurus OB association 
(Sco-Cen) contains one of the largest nearby populations of newly-formed stars 
\citep[$d\sim145$~pc, $\tau\sim10$~Myr,][]{pm08}.
Since it extends across a large area of sky ($\sim100$~deg$^2$),
a census of its members requires wide-field imaging data.
The extinction in Upper Sco is low enough ($A_V<3$~mag) that it has been
possible to perform wide-field surveys of the association across a broad
range of wavelengths \citep[e.g.,][]{pre98,lod06,sle06,riz15,pec16}.
\citet{luh18} compiled stars within the confines of Upper Sco
that have spectral classifications and evidence of youth and whose available
measurements of proper motion and parallax are consistent with membership,
resulting in a catalog of more than 1600 proposed members. That study also
presented $\sim1200$ additional candidate members that lacked spectroscopy.

\citet{luh18} utilized data from several surveys,
including the first data release (DR1) from the 
{\it Gaia} mission \citep{per01,deb12,gaia16b}, which provided 
an all-sky catalog of high-precision proper motions and
parallaxes for stars with $G\lesssim12$~mag \citep{gaia16a}.
Since \citet{luh18}, the second data release of {\it Gaia} (DR2) has been
issued, which has extended measurements of those parameters down to much
fainter magnitudes \citep[$G\sim20$~mag,][]{bro18}.
Data from DR1 and DR2 have been applied extensively to young clusters
and association for the measurement of the distances and internal kinematics
of known members \citep{dzi18,gal18,roc18,wri18,wri19,kuh19} and for
the identification of candidate members 
\citep{coo17,oh17,bec18,wil18,gag18c,gag18d,fah18,luh18tau,ros18,her19,gal20}.
The DR2 magnitude limit for parallaxes and proper motion corresponds to
especially low masses in Upper Sco ($\sim0.04$~$M_\odot$) due to the
combination of its proximity, low extinction, and youth.
Indeed, DR2 has been used to identify candidate members of Upper Sco and
other populations in Sco-Cen and to examine spatial and kinematic substructure
among those candidates \citep{gol18,can19,dam19}.

We have sought to further improve the census of Upper Sco using data from
{\it Gaia} DR2 and new spectroscopy of candidate members.
We begin by using {\it Gaia} data to characterize the kinematics
of the stellar populations within Sco-Cen and to define kinematic criteria for
membership in Upper Sco (Section~\ref{sec:criteria}). Those criteria are
incorporated into the selection of candidate members of Upper Sco from
\citet{luh18} and spectral classifications are presented for a
few hundred candidates (Section~\ref{sec:search}).
We compile a catalog of stars toward Upper Sco that exhibit spectroscopic
evidence of youth and assess their membership in the association
(Section~\ref{sec:classify}). Using the adopted members and
the most promising remaining candidates that lack spectra (i.e., those 
with {\it Gaia} astrometry),
we characterize the stellar population of Upper Sco in terms of its
initial mass function (IMF), disk fraction, and median age
(Section~\ref{sec:pop}).
An appendix also includes comments on individual sources,
extinction coefficients in the {\it Gaia} bands based
on the extinction curve from \citet{sch16}, and new estimates for the typical
intrinsic colors of young stellar photospheres in optical and infrared (IR)
bands.

\section{Kinematic Membership Criteria for Upper Sco}
\label{sec:criteria}

\subsection{{\it Gaia} DR2}
\label{sec:dr2}

To improve upon the recent census of Upper Sco from \citet{luh18}, we
make use of new data from {\it Gaia} DR2.
They include photometry in bands at 3300--10500~\AA\ ($G$), 
3300--6800~\AA\ ($G_{\rm BP}$), and 6300-10500~\AA\ ($G_{\rm RP}$),
proper motions and parallaxes for most stars down to $G\sim20$~mag,
and radial velocities primarily for $G\sim4$--12~mag.
The additional contents of DR2 are summarized by \citet{bro18}.

\begin{figure}
\epsscale{0.6}
\plotone{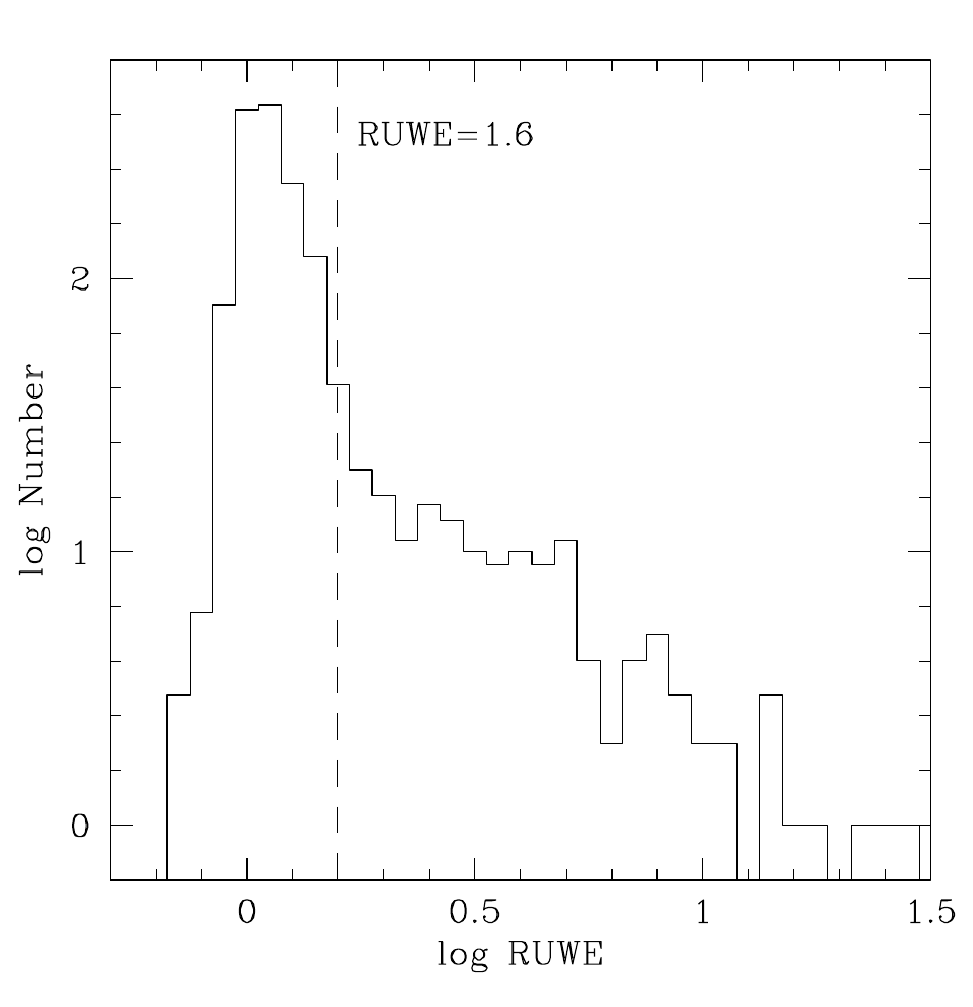}
\caption{
Distribution of log(RUWE) for stars adopted as members of the Upper Sco
association by \citet{luh18} that have parallax measurements from {\it Gaia}
DR2 (histogram). RUWE characterizes the quality of the astrometric fit for
a given star \citep{lin18}. We treat astrometry for stars with RUWE$<$1.6
as reliable.
}
\label{fig:ruwe}
\end{figure}

DR2 contains parameters that can be used to identify stars
with poor astrometric fits, and hence potentially unreliable astrometry
\citep{lin18b}. After DR2, \citet{lin18} defined a new parameter to serve as
a better indicator of the goodness of fit, the renormalized unit weight
error (RUWE). The distribution of RUWE in DR2 exhibited a maximum
centered near unity (well-behaved fits) and a long tail at values higher
than $\sim$1.4. Therefore, \citet{lin18} suggested that RUWE$\lesssim$1.4 
could be used as a threshold for reliable astrometry.
\citet{esp19} examined the applicability of that threshold to the Taurus
star-forming region, and found that a break does appear near
that value in the distribution of RUWE for the known members.
The fraction of members with discrepant astrometry increased with larger
values of RUWE above 1.4, although many members above that threshold
had astrometry that was consistent with membership, and thus did not appear
to have large errors.
We have repeated that exercise for Upper Sco in Figure~\ref{fig:ruwe}, where
we plot the distribution of log(RUWE) for the members adopted by \citet{luh18}.
The population of well-behaved fits near unity is slightly broader in
Upper Sco than in Taurus, so we adopt a larger threshold (RUWE$<$1.6)
when selecting reliable astrometry for our analysis.
As with Taurus, many of the previously adopted members of Upper Sco 
with high values of RUWE have astrometry that is consistent with membership.

\subsection{Kinematics in Sco-Cen}
\label{sec:kin}

\begin{figure}
\epsscale{1.2}
\plotone{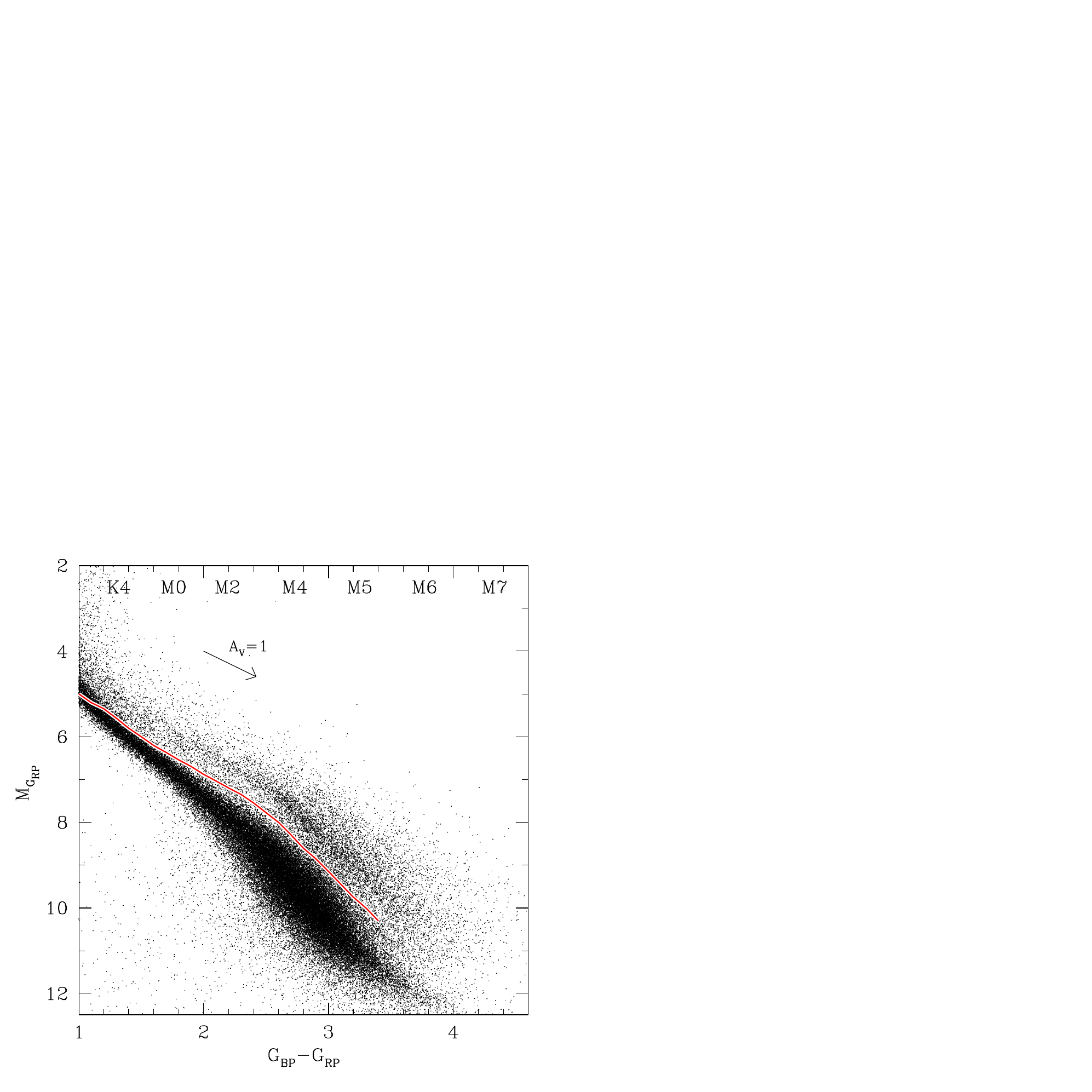}
\caption{
$M_{G_{\rm RP}}$ versus $G_{\rm BP}-G_{\rm RP}$ for stars within the
boundary of Sco-Cen defined by \citet{dez99} that have
$\pi>5$~mas, $\pi/\sigma\geq20$, and RUWE$<$1.6 in {\it Gaia} DR2.
We have selected a sample of candidate young low-mass stars based on 
colors of $G_{\rm BP}-G_{\rm RP}$=1.4--3.4~mag ($\sim$0.15--1~$M_\odot$)
and positions above
the single-star sequence for the Tuc-Hor association \citep[45~Myr,][]{bel15}
(solid line). 
We have marked a reddening vector for the extinction curve
from \citet{sch16} and the spectral types that
correspond to these colors for young stars (see Appendix).
}
\label{fig:br}
\end{figure}

As often done in surveys for new members of clusters and associations,
\citet{luh18} used previously known young stars in the direction of Upper Sco
to define photometric and kinematic criteria for the selection of new
candidate members. However, because Upper Sco is one of several populations
within the Sco-Cen complex, some of which overlap on the sky,
we examine the kinematics of the entire complex in order to identify those
that apply to Upper Sco.
To reveal the kinematics of the various populations in Sco-Cen, we begin by
identifying a sample of candidate young stars using a color-magnitude diagram
(CMD). In Figure~\ref{fig:br}, we plot 
$M_{G_{\rm RP}}$ versus $G_{\rm BP}-G_{\rm RP}$ for stars from DR2
that are within the boundary of Sco-Cen from \citet{dez99}, are 
close enough to be members of the complex ($\pi>5$~mas), and have 
small relative errors in parallax
($\pi/\sigma\geq20$) and reliable astrometry (RUWE$<$1.6).  
The threshold in $\pi/\sigma$ was chosen to be high enough
that measurement errors in parallax and proper motion are a negligible
component of the spread in those parameters in Sco-Cen while low enough
to produce a sufficiently large sample of stars for fully sampling
the kinematics of Sco-Cen.
We have selected a sample of candidate young low-mass stars
based on colors between $G_{\rm BP}-G_{\rm RP}$=1.4--3.4~mag ($\sim$K5--M5,
$\sim$0.15--1~$M_\odot$) and absolute magnitudes above the single-star sequence
for the Tuc-Hor association \citep[45~Myr,][]{bel15}, which is older
than the oldest isochronal ages for known members of Sco-Cen 
\citep[$\lesssim30$~Myr,][]{pec12,pec16}.

\begin{figure}
\epsscale{1}
\plotone{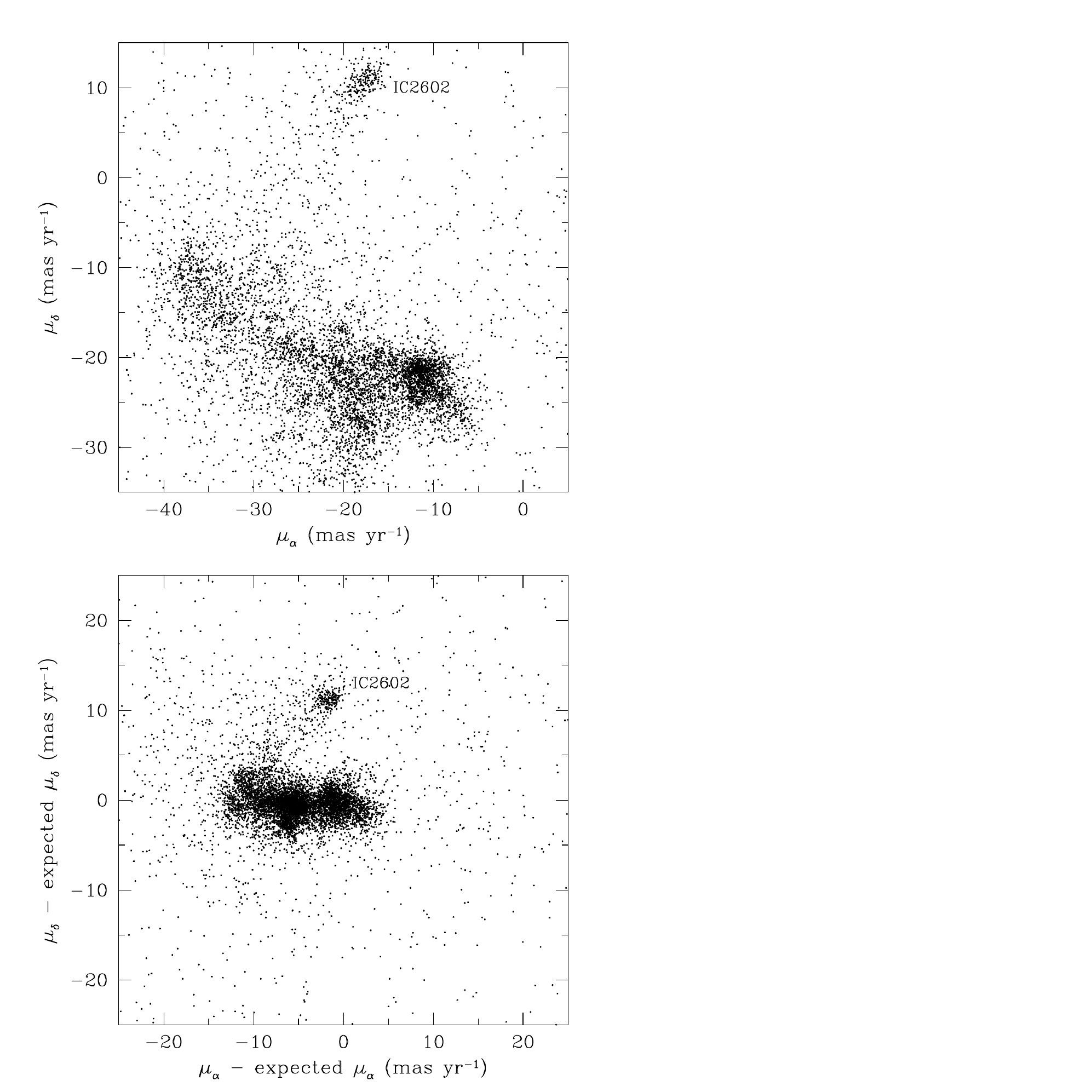}
\caption{
Proper motions and proper motion offsets for candidate young stars within
the boundaries of Sco-Cen selected from Figure~\ref{fig:br}.
The offsets are computed relative to the proper motions expected
for the positions and parallaxes assuming the median space velocity of Upper
Sco members ($U, V, W = -5, -16, -7$~km~s$^{-1}$, Section~\ref{sec:classify}). 
}
\label{fig:pm}
\end{figure}

In Figure~\ref{fig:pm}, we have plotted the proper motions
($\mu_{\alpha,\delta}$) and proper motion offsets 
($\Delta\mu_{\alpha,\delta}$)
from {\it Gaia} DR2 for the candidate young stars identified with
the CMD in Figure~\ref{fig:br}. 
The proper motion offset is defined as the difference
between the observed proper motion for a given star and the motion expected at
the star's celestial coordinates and parallax if it had
the median space velocity of members of Upper Sco 
from \citet{luh18} rounded to the nearest integers
($U, V, W = -5, -16, -7$~km~s$^{-1}$, see also Section~\ref{sec:classify}).
Stars with the same space velocities but different celestial coordinates
exhibit different proper motions. The proper motion offsets serve to reduce
those projection effects, which is why the candidate young stars
in Figure~\ref{fig:pm} are more tightly clustered in those parameters.
The smaller concentration of proper motion offsets corresponds to the open
cluster IC~2602, which has an age of 46~Myr \citep{dob10}. 
The larger group of offsets contains the populations in Sco-Cen.

\begin{figure}
\epsscale{1}
\plotone{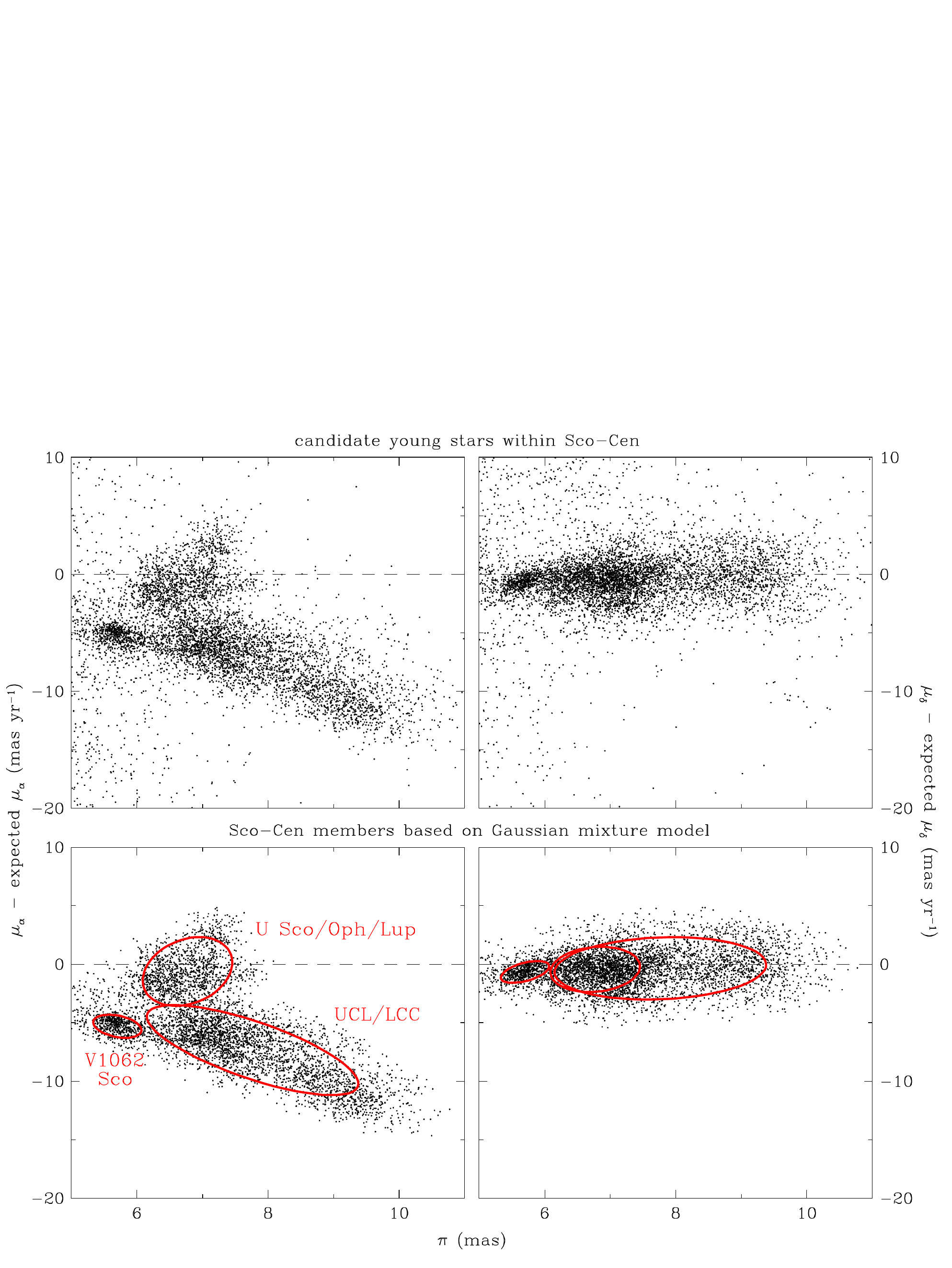}
\caption{
Proper motion offsets versus parallax for candidate young stars within
the boundaries of Sco-Cen selected from Figure~\ref{fig:br} (top).
A Gaussian mixture model has been applied to these data to estimate
probabilities of membership in Sco-Cen and the field population.
The stars that have $>90$\% probabilities of membership in Sco-Cen are shown
(bottom). The three Sco-Cen components in the model are represented by the 
ellipses (2~$\sigma$).
}
\label{fig:pp4}
\end{figure}

To include parallax in our kinematic analysis of Sco-Cen,
we have plotted $\Delta\mu_{\alpha}$ and $\Delta\mu_{\delta}$ versus
parallax in the top row of Figure~\ref{fig:pp4} for the candidate young
stars from Figure~\ref{fig:br}.
These data exhibit a rich, complex distribution and a sparse, widely
scattered distribution, which correspond to Sco-Cen and field stars,
respectively. Many of these candidate members of Sco-Cen have been
previously identified with {\it Gaia} DR2 \citep{gol18,can19,dam19}.
We have estimated probabilities of membership in Sco-Cen and
the field population by applying a Gaussian mixture model to the 
parallaxes and proper motion offsets using the {\tt mclust} package in
R \citep{Rcore13,mclust}.
The mixture model included a noise component
for the field stars, which was initialized using the nearest neighbor cleaning 
method from \citet{bye98} via the {\tt NNclean} function in the {\tt prabcus}
library of R. Fits were performed with models in which the number
of components ranged from 1 to $\sim20$.
We have adopted the model with the smallest number of components that
is effective at discriminating between the field stars and members of Sco-Cen
based on the distributions of membership probabilities, which 
corresponds to the model with three components for Sco-Cen.
In that model, most stars ($\sim$95\%) have a high probability 
($>$90\%) of membership in either Sco-Cen or the field population.
That fraction does not increase significantly for models with more than
three components.
We note that the Gaussian mixture model applied by {\tt mclust} does not
make use of measurement errors. However, the errors in parallaxes and
proper motion offsets for the sample in question are much smaller than
the ranges of values in Sco-Cen, so the identification of populations
and the estimates of membership probabilities should not depend
significantly on the inclusion of those errors.

\begin{figure}
\epsscale{1}
\plotone{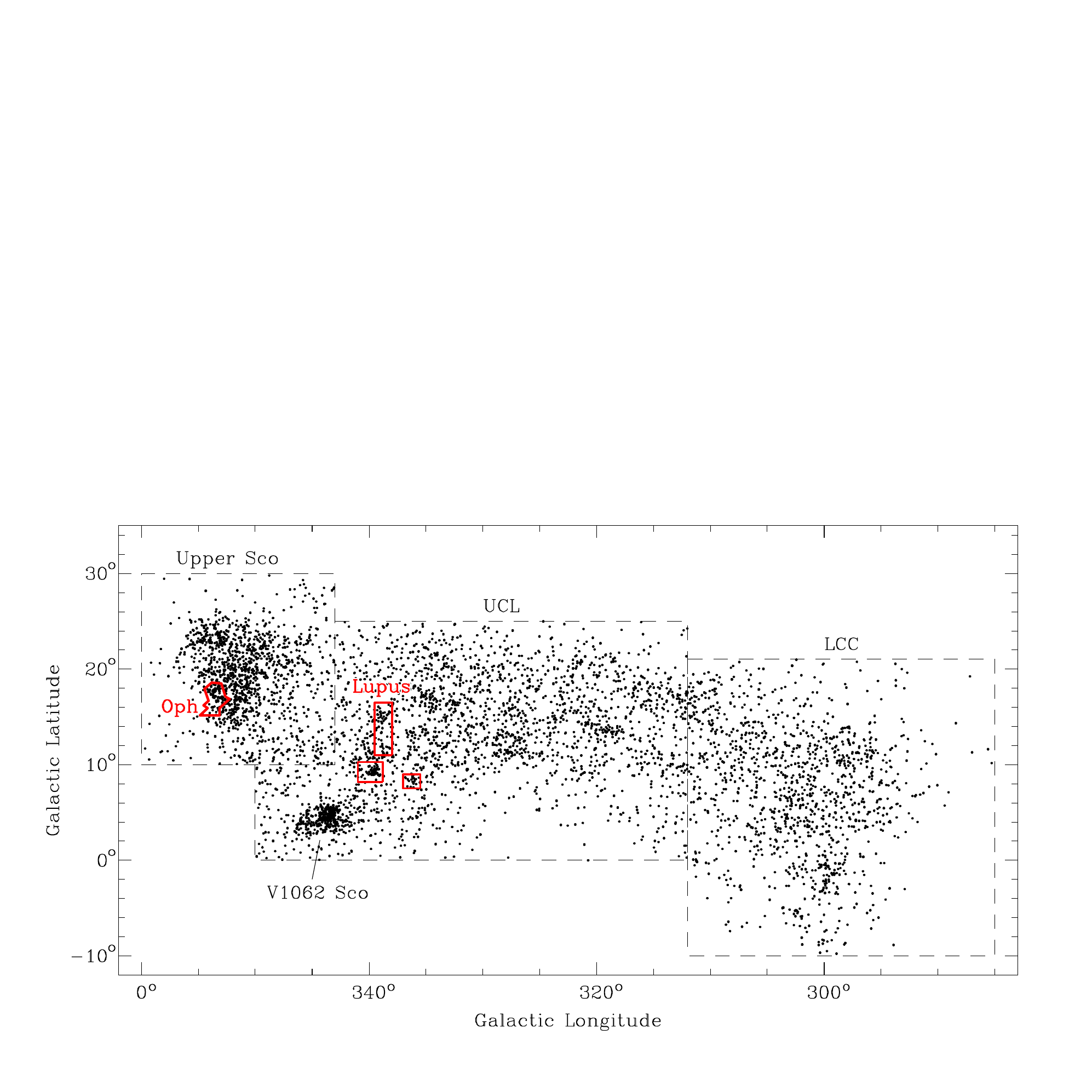}
\caption{
Spatial distribution of the candidate members of Sco-Cen with
$>90$\% membership probabilities from Figure~\ref{fig:pp4}.
The boundaries from \citet{dez99} for the subgroups
are indicated (dashed lines). We also have marked the boundary
between Upper Sco and Ophiuchus from \citet{esp18}
and rectangles that encompass clouds 1--4 in Lupus (solid red lines).
}
\label{fig:mapsc}
\end{figure}

The parallaxes and proper motion offsets for the 5353 stars with $>$90\%
probabilities of membership in Sco-Cen are plotted in the bottom row of
Figure~\ref{fig:pp4}. We have included an ellipse (2~$\sigma$) for each
of the three components in the model.
A map of those stars is presented in Figure~\ref{fig:mapsc}.
We have marked the boundaries for the three subgroups of Sco-Cen from 
\citet{dez99}, which consist of Upper Sco, Upper Centaurus-Lupus (UCL) and
Lower Centaurus-Crux (LCC). We also have included the boundary between Upper
Sco and the Ophiuchus star-forming region that was adopted by \citet{esp18}
and rectangles that encompass clouds 1--4 in the Lupus star-forming region.
In Figure~\ref{fig:mapsc}, the densest concentrations of stars are found in 
Upper Sco and in a cluster associated with V1062~Sco \citep{ros18,dam19}. 
The remaining stars are more widely scattered across the subgroups of Sco-Cen. 
The three components of our adopted mixture model 
correspond primarily to Upper Sco/Ophiuchus/Lupus, the V1062~Sco cluster, 
and UCL/LCC, as indicated in Figure~\ref{fig:pp4}. 

\begin{figure}
\epsscale{1}
\plotone{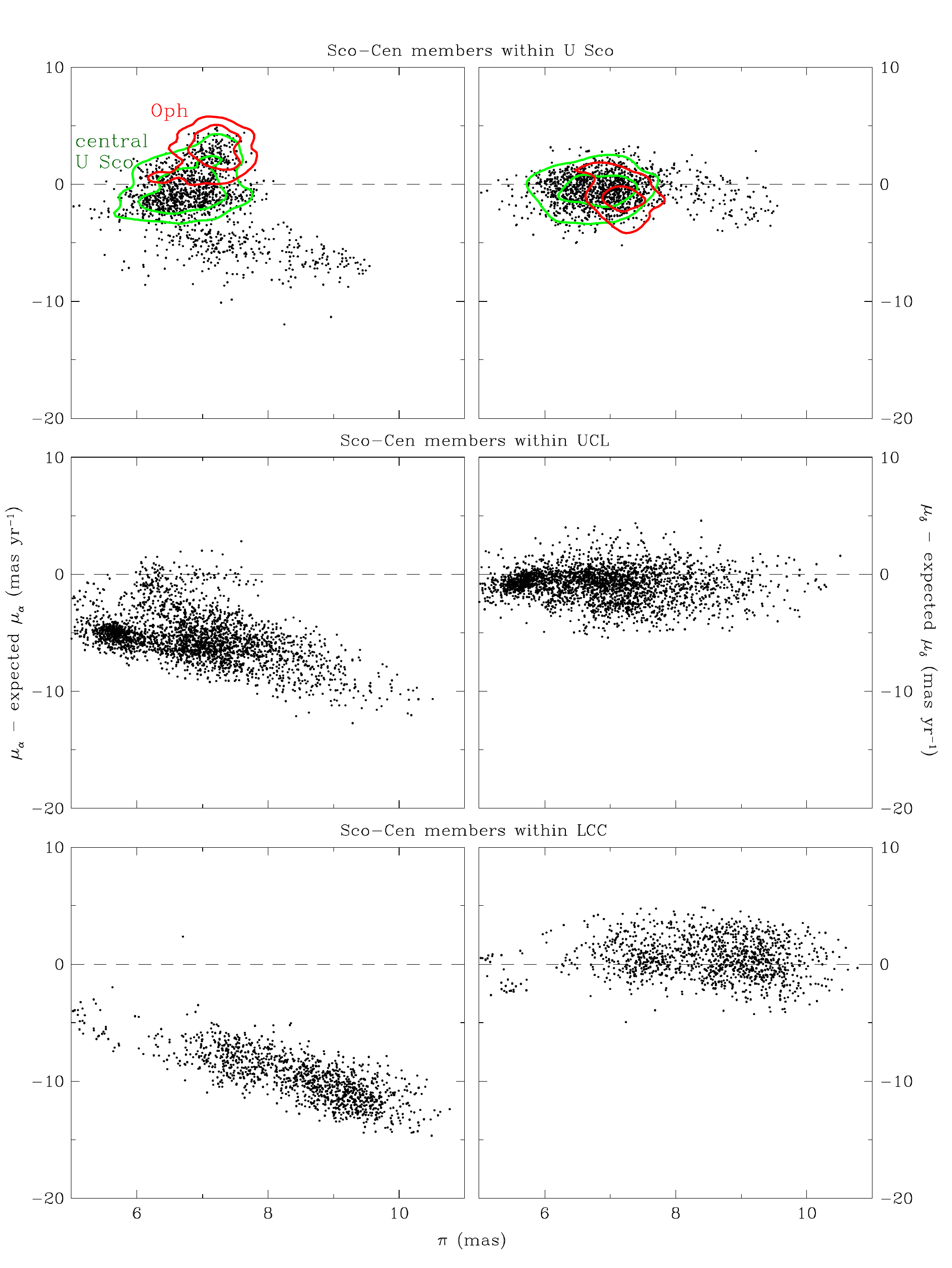}
\caption{
Proper motion offsets versus parallax for the candidate members of Sco-Cen with
$>90$\% membership probabilities (Figure~\ref{fig:pp4})
that are within the boundaries of Upper Sco (top),
UCL (middle), and LCC (bottom) from \citet{dez99}.
We have included contours for density maps of the data within
the central concentration in Upper Sco and within the boundary of Ophiuchus from
\citet{esp18} (see Fig.~\ref{fig:mapsc}).  For each of those populations,
the contours are plotted at 10\% and 50\% of the maximum density.
}
\label{fig:pp}
\end{figure}

The variation of kinematics across Sco-Cen is illustrated in
Figure~\ref{fig:pp}, which shows the parallaxes and proper motion offsets
for the candidate members from Figure~\ref{fig:mapsc} that within 
the boundaries of each of the three subgroups.
The broadest population in parallax from our mixture model is concentrated
within the boundaries of UCL and LCC, but it extends across Upper Sco as well.
That population exhibits a shift in parallax and $\Delta\mu_{\alpha}$ between
UCL and LCC. To better isolate the kinematics of Upper Sco and Ophiuchus, we 
have used the functions {\tt kde2d} and {\tt bandwidth.nrd}
from the {\tt MASS} package in R \citep{mass}
to calculate density maps in $(\pi,\Delta\mu_{\alpha})$ and
$(\pi,\Delta\mu_{\delta})$ for the Sco-Cen members within the boundary of
Ophiuchus from \citet{esp18} and for members outside of that boundary and
within the triangular area in which most Upper Sco members are concentrated
(Fig.~\ref{fig:mapsc}). The latter region is referred to as the ``central
concentration" of Upper Sco henceforth in this study.
Contours for the resulting density maps are plotted in Figure~\ref{fig:pp},
which show a shift in the kinematics between the two populations
\citep[see also][]{dam19}. For the purpose of assessing membership in
Upper Sco and Ophiuchus in Section~\ref{sec:memclass},
we have adopted thresholds in parallax, $\Delta\mu_{\alpha}$, and
$\Delta\mu_{\delta}$ that approximate the 10\% contours for the two
populations in Figure~\ref{fig:pp}.
In a similar way, we have also defined thresholds for membership in
UCL/LCC that follow the outer contours of a density map for the sum
of the UCL and LCC samples in Figure~\ref{fig:pp}.
We note that young stars associated with the Lupus
clouds \citep[][references therein]{com08} are similar to Upper Sco in
their proper motion offsets but are slightly more distant with a
median parallax of $\sim$6.3~mas.

Since our Gaussian mixture model was applied only to parallax and proper motion
offsets, it is possible for members of a given population to be widely separated
on the sky. For instance, the diagrams in Figure~\ref{fig:pp} for
LCC contain several stars that have kinematics similar to the V1062~Sco cluster,
which is more than $30\arcdeg$ from LCC. Given the compact nature of V1062~Sco, 
it seems likely that those stars are unrelated to V1062~Sco or Sco-Cen.
We have investigated mixture models that include celestial coordinates,
but the resulting populations were less satisfactory because of the non-Gaussian
distributions of positions of young stars across UCL and LCC.

We have compared the sample of 5353 stars with $>$90\% membership probabilities
in Sco-Cen from our Gaussian mixture model to the sample of 10,839 candidate
members identified with {\it Gaia} DR2 by \citet{dam19}.
We find that 5295 of our candidates (99\%) are also in the sample from
\citet{dam19}. Among the 5544 candidates from \citet{dam19}
that are absent from our sample, 4713 are candidate members of IC~2602 (which
we ignored) or are outside of the ranges of values that we considered
for celestial coordinates, $G_{\rm BP}-G_{\rm RP}$, or RUWE.
An additional 200 of those 5544 candidates appear below the boundary
in Figure~\ref{fig:br} that we used for identifying young stars.
The remaining 631 candidates from \citet{dam19} that are absent from our
sample have low probabilities
of membership in the three components of our Gaussian mixture model that
correspond to Sco-Cen. Those stars have proper motion offsets and
parallaxes in Figure~\ref{fig:pp} that differ significantly from the
three Sco-Cen components, so they were are assigned to the noise component
(field stars) by our model.

\section{Search for New Members of Upper Sco}
\label{sec:search}

\subsection{Application of {\it Gaia} DR2}
\label{sec:app}

Using CMDs, proper motions, and parallaxes (when available),
\citet{luh18} identified candidate members of Upper Sco 
within the area from $\alpha=15^{\rm h}35^{\rm m}$ to $16^{\rm h}45^{\rm m}$ and
$\delta=-30$ to $-16\arcdeg$ and outside of the boundary of Ophiuchus
defined by \citet{esp18}.
They considered astrometry and photometry from the Point Source Catalog of 
the Two Micron All Sky Survey \citep[2MASS,][]{skr06}, the third data 
release of the Deep Near-Infrared Survey of the Southern Sky 
\citep{epc99}, the science verification release and data release 10 
(DR10) of the United Kingdom Infrared Telescope Infrared Deep Sky Survey
\citep[UKIDSS,][]{law07}, the AllWISE Source Catalog of
the {\it Wide-field Infrared Survey Explorer}
\citep[{\it WISE},][]{wri10}, the first data release of the 3$\pi$ survey
from Pan-STARRS1 \citep[PS1,][]{kai02,kai10,cha16,fle16}, 
DR1 from {\it Gaia} \citep{gaia16a}, and the fifth data release of 
the Visible and Infrared Survey Telescope for Astronomy (VISTA) 
Hemisphere Survey \citep[VHS,][]{mcm13}.
\citet{luh18} presented a catalog of 1196 candidate members of Upper Sco
that were identified with those data and that lacked spectroscopic
confirmation of their youth.
In this study, we have updated the VHS photometry in our candidate selection
with data from the sixth data release (DR6).
In terms of $JHK$ photometry, UKIDSS and VHS exhibit a systematic offset
from 2MASS that varies with color, and hence spectral type, due
to differences in the transmission profiles of their filters \citep{hod09}.
We have applied offsets to the UKIDSS and VHS data to align them with
the 2MASS data, on average, for the known M-type members of Upper Sco
(most members at earlier types are saturated in UKIDSS and VHS),
which corresponds to adding 0.07, $-0.04$, and 0.03~mag to $J$, $H$, and $K$
from UKIDSS and adding 0.07 and $-0.03$~mag to $J$ and $K$ from VHS,
respectively. These offsets were not included in the analysis from
\citet{luh18}.

To further refine the sample of candidates from \citet{luh18},
we can employ the kinematic criteria for membership from the previous
section (Figure~\ref{fig:pp}) using {\it Gaia} DR2.
We identified those criteria by considering
stars with sufficiently precise astrometry ($\pi/\sigma\geq20$) that
we could accurately characterize the kinematics of Upper Sco.
We now apply those criteria to a broader range of parallax errors
($\pi/\sigma\geq10$) when identifying candidate members of Upper Sco.
Data from DR2 are available for 1095 of the 1196 candidates, 972 of which
have $\pi/\sigma\geq10$.
Among the stars with $\pi/\sigma\geq10$ and reliable astrometry as indicated
by RUWE$<$1.6 (Section~\ref{sec:dr2}),
we have rejected those that do not overlap 
(at $1\sigma$)\footnote{The errors in proper motion offsets
include both the errors in the proper motions and the errors in the expected
motions due to the parallax measurements.}
with the ranges of parallax and proper motion offsets for the central
central concentration of Upper Sco in Figure~\ref{fig:pp}, which results
in the rejection of 370
candidates. For stars with RUWE$>$1.6, agreement with those criteria is taken as
additional evidence supporting candidacy and disagreement is ignored
and does not affect candidacy.
For stars that lack precise parallaxes but that have single-epoch
positions from DR2, we have replaced the 2MASS/{\it Gaia} DR1 proper motions
from \citet{luh18} with new measurements using 2MASS and the DR2 positions.
Those updated proper motions lead to the rejection of four candidates.
Following our new adjustments to the UKIDSS and VHS photometry described
earlier in this section, 20 additional candidates from \citet{luh18}
no longer satisfy the selection criteria in the CMDs from that study.
Thus, a total of 394 candidates from \citet{luh18} are now rejected.

Since many of the proper motion measurements employed by \citet{luh18} are
superseded by more accurate data from {\it Gaia} DR2 or 2MASS/{\it Gaia} DR2,
some stars that were rejected in that study now qualify as candidate members.
In addition, stars with $\pi/\sigma\geq10$ and RUWE$<$1.6
are adopted as candidates if they satisfy the kinematic criteria for
membership from the previous section, appear above the single-star
sequence for the Tuc-Hor association in a diagram of $M_{G_{\rm RP}}$ versus
$G_{\rm BP}-G_{\rm RP}$ (Figure~\ref{fig:br}), and are not previously
identified members or candidates.
These candidates consist of stars that were rejected by a few of the CMDs
from \citet{luh18} or lacked sufficient photometry to be selected
as candidates in that study because they are close companions or 
bright stars that are absent from the 2MASS Point Source Catalog.  
We also selected stars that have $\pi/\sigma\geq10$, satisfy the kinematic
criteria in Figure~\ref{fig:pp}, lack photometry in $G_{\rm BP}$ or
$G_{\rm RP}$, and are within $5\arcsec$ of a candidate or known member.
Finally, some stars that were rejected by CMDs in \citet{luh18} now
satisfy those diagrams because of the new offsets applied to the photometry
from UKIDSS and VHS. In total, we have identified 155 candidate members of
Upper Sco that were not in the catalog of candidates from \citet{luh18}.

We have rejected 394 of the 1196 candidate members of Upper Sco
from \citet{luh18} and have added 155 new candidates, resulting in
a sample of 957 candidates.
In the next section, we classify the youth of 382 of those 957 candidates
using new spectra that we have collected.
Five additional candidates (2MASS J16232181$-$2424577, 
2MASS J16250449$-$2509114, 2MASS J16260214$-$2718141,
2MASS J15562344$-$2541056, 2MASS J16212830$-$2529558)
exhibit evidence of youth in previous spectroscopy \citep{mar98a,tuc15,chi20}.
There remain 567 candidates that lack sufficient spectroscopic data to
assess their youth, which are listed in Table~\ref{tab:cand} with
their available spectral classifications, astrometry and
photometry from {\it Gaia} DR2, and IR photometry.
The central concentration of Upper Sco (Figure~\ref{fig:mapsc}) contains
$\sim200$ of these candidates.

\citet{esp18} presented 30 candidate disk-bearing members of Upper Sco that
lacked spectroscopic confirmation of their youth, 26 of which satisfy
the selection criteria described in this section. Three of the remaining
candidates are rejected based on their astrometry from {\it Gaia} DR2
(AllWISE J162204.36$-$240814.5, AllWISE J155233.92$-$265112.5,
AllWISE J155125.62$-$270743.4) and one candidate (2MASS J16214330$-$2345490)
no longer satisfies one of the CMDs from \citet{luh18} due to our new 
adjustments to the UKIDSS photometry.

Among the 567 candidates in Table~\ref{tab:cand}, 28 are within $5\arcsec$
of known young stars, and thus are candidate companions. All of these
candidates have parallaxes and proper motions from {\it Gaia} DR2 that
support membership in Upper Sco.
For each candidate, we include in Table~\ref{tab:cand} the
separation and {\it Gaia} designation of the nearest known young star within
$5\arcsec$. In addition, there are 17 pairs of candidates in
Table~\ref{tab:cand} that have separations less than $5\arcsec$, which may
also comprise binary systems.

Six of the candidates in Table~\ref{tab:cand} are bright enough that they should
be easily detected by {\it Gaia} ($i<15$~mag, $J<13$~mag) but are absent from
DR2. Forty-four candidates have {\it Gaia} counterparts but lack parallax
measurements, 34 of which have RUWE$>$1.6, indicating poor astrometric fits.

We estimated the completeness limits of the $H$ and $K$ photometry
from UKIDSS to be 17.4 and 17.0~mag, respectively, based on the magnitudes
at which the logarithm of the number of sources as a function of magnitudes
departs from a linear slope and begins to turn over.
When those limits are combined with the sequence of known members
of Upper Sco in spectral type versus $H$ and $K$ \citep{luh18},
they imply a completeness limit of $\sim$L0 ($\sim0.01$~$M_\odot$)
for our survey for candidate members.

As mentioned at the start of this section, the survey for candidate members
of Upper Sco from \citet{luh18} and in this work applies to an area
outside of the boundary of Ophiuchus adopted by \citet{esp18}.
\citet{esp20} presents a catalog of candidate members
of Upper Sco and Ophiuchus within that boundary.

\subsection{Spectroscopy of Candidates}
\label{sec:spec}

\citet{luh18} and \citet{esp18} obtained optical and IR spectra
of several hundred candidate members of Upper Sco to measure their spectral
types and check for signatures of youth that would support their membership.
We have continued that work through spectroscopy of additional candidates.
Our observations were performed with 
the 4~m Blanco telescope at the Cerro Tololo Inter-American
Observatory (CTIO), the Magellan Baade telescope at Las Campanas Observatory,
the NASA Infrared Telescope Facility (IRTF), the Bok Reflector at Steward
Observatory, and the Gemini North and South telescopes.
The instrument configurations are summarized in Table~\ref{tab:log}.
We collected spectra of 430 objects, which are listed in Table~\ref{tab:spec}
with the instrument and date of each observation.
We also have made use of a publicly available spectrum of one additional
candidate, 2MASS J16173081$-$2411411, that was taken at Gemini North through
program GN-2018A-Q-115 (M. Liu).
When selecting these targets, we gave highest priority to candidates within
the central concentration in Upper Sco (Figure~\ref{fig:mapsc}).
One of the targets is a previously known member that was mistakenly
observed with spectroscopy ({\it Gaia} 6049502122647926912).
Most of the targets (401) satisfy our latest criteria for candidate
members from the previous section. Among the remaining 29 targets, 
21 were candidates from \citet{luh18} that were observed before our
selection criteria involving {\it Gaia} DR2 were finalized;
one was observed because of IR excess emission and a previous classification
that was uncertain \citep[{\it Gaia} 6243223873758073728,][]{esp18};
two were classified as galaxies by \citet{luh18} and \citet{esp18}
based on emission lines and featureless continua in IR spectra,
but {\it Gaia} DR2 data supported membership, so they were observed again
in this work ({\it Gaia} 6242434969863194880 and
6243210920137332096); and five were observed because they are located
within a few arcseconds of candidate members ({\it Gaia} 6045893074511675008,
6051816251024722176, 6049264044017673856, and 6045405136172296192 and
2MASS J16100274$-$2344400).

\begin{figure}
\epsscale{1}
\plotone{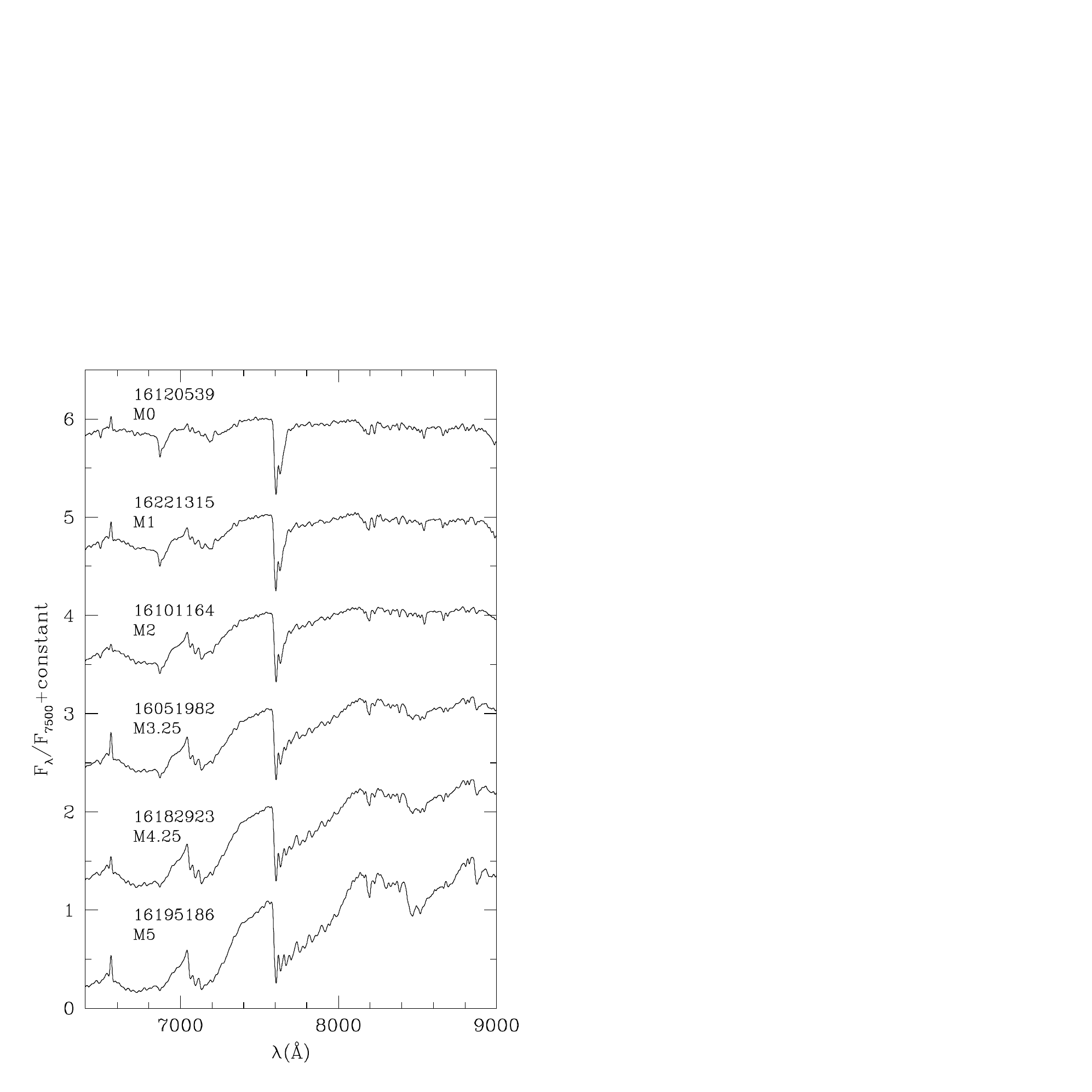}
\caption{
Examples of optical spectra of members of the Upper Sco association.
These data are displayed at a resolution of 13~\AA.
The data used to create this figure are available.
}
\label{fig:op}
\end{figure}

\begin{figure}
\epsscale{1}
\plotone{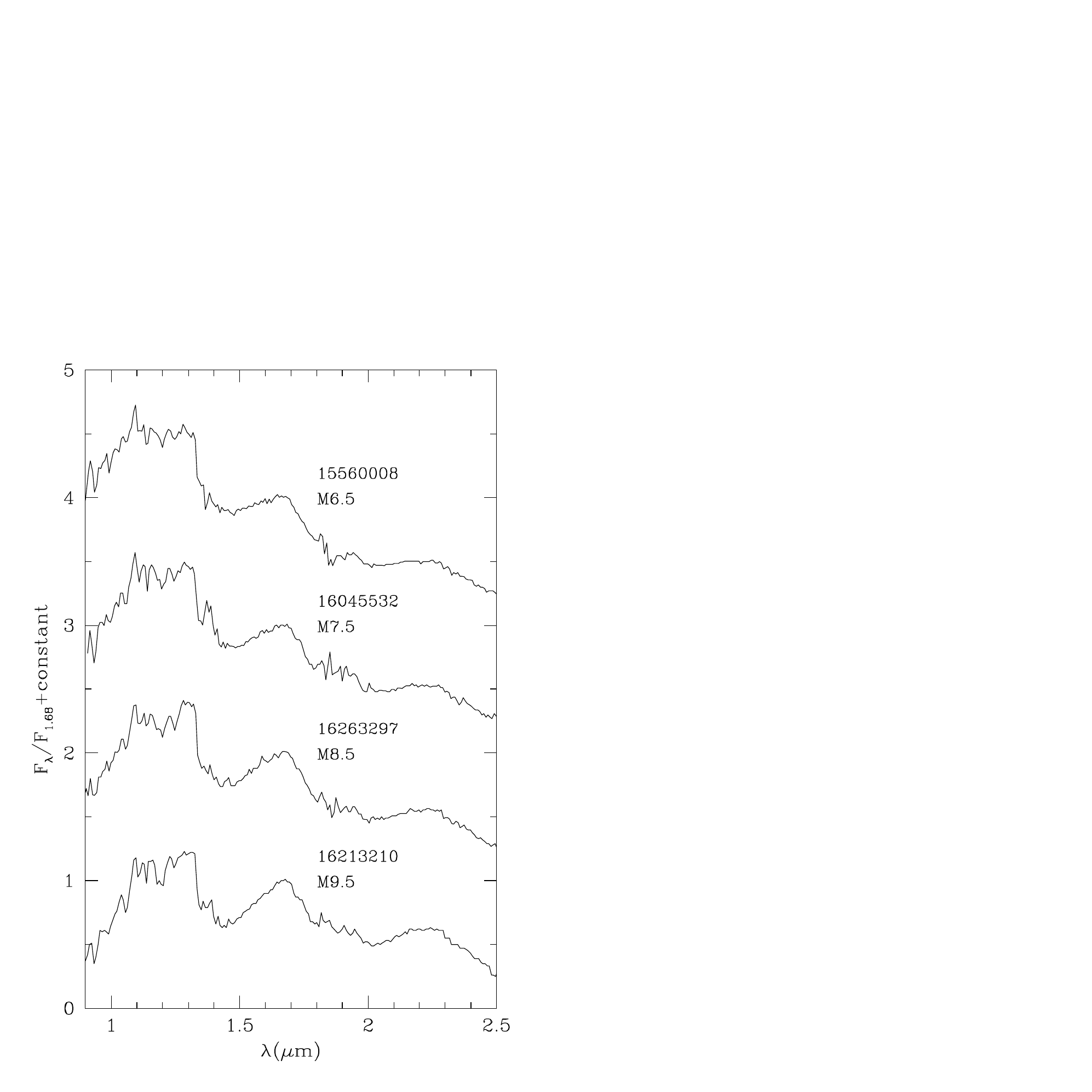}
\caption{
Examples of near-IR spectra of M-type members of the Upper Sco association.
They have been dereddened to match the slopes of the young standards from
\citet{luh17} and are displayed at a resolution of $R=200$.
The data used to create this figure are available.
}
\label{fig:ir}
\end{figure}

We reduced the SpeX data with the Spextool package \citep{cus04}, which
included correction of telluric absorption \citep{vac03}.
The spectra from the other instruments were reduced using routines
within IRAF. We present examples of the reduced optical and IR
spectra in Figures~\ref{fig:op} and \ref{fig:ir}, respectively.
The reduced spectra are provided in electronic files that accompany
those figures.
The optical spectra typically have signal-to-noise ratios (S/N's)
of $\sim50$ at 7400--7500~\AA.
Most of the IR spectra exhibit S/N$\sim50$--100 in the $H$ and $K$ bands
when binned to the lowest resolution among those data ($\sim$150).

We have analyzed our spectra with the same methods employed in our recent
surveys for new members of Upper Sco \citep{esp18,luh18}. 
In brief, we examined the spectra for evidence that the targets are
young enough to be members of Upper Sco ($\sim10$~Myr) based on
Li absorption at 6707~\AA\ and gravity-sensitive features like the Na doublet
near 8190~\AA\ and the near-IR H$_2$O absorption bands.
We also measured spectral types via comparison to standard spectra 
for field dwarfs \citep{kir91,kir97,hen94,cus05,ray09}
and members of star-forming regions \citep{luh97,luh99,luh17}.
Our measurements of spectral types, equivalent widths of Li, and
assessments of youth are listed in Table~\ref{tab:spec}.
Most of the targets (376/431) exhibit evidence of youth.
Comments on the classifications of individual sources are provided in
the Appendix.

\section{Classifying the Membership of Young Stars Toward Upper Sco}
\label{sec:classify}

\citet{luh18} compiled a catalog of 1631 stars that exhibited evidence
of membership in Upper Sco and that are located outside of the boundary
of Ophiuchus defined by \citet{esp18}.
We have updated the analysis that produced that catalog to include the
new young stars from this work, a few young stars toward Upper Sco from
previous studies that were absent in \citet{luh18}, and constraints
on membership provided by astrometry from {\it Gaia} DR2 
(Section~\ref{sec:kin}).

We have compiled all stars within the survey field considered by
\citet{luh18} (see Section~\ref{sec:app}) that exhibit evidence of youth
from previous studies and this work. As done in that study, we include
a few stars outside of that field that have been widely adopted as
members of Upper Sco.
Early-type stars lack spectroscopic signatures of youth, so we have
considered all A and B stars that do not have parallax or proper motions
that are highly discrepant ($>2$~$\sigma$) from the membership criteria in
Section~\ref{sec:app}. The resulting sample contains 2020 objects
and is presented in Table~\ref{tab:mem}.
We have tabulated the available measurements of spectral types,
our adopted types, our extinction estimates, proper motions and parallaxes
from {\it Gaia} DR2, radial velocities from {\it Gaia} and previous studies,
our calculations of $UVW$ space velocities based on the astrometry from
{\it Gaia} and the radial velocities, photometry from {\it Gaia} DR2,
$JHK_s$ photometry from various
sources, our kinematic classifications of membership, and a flag indicating
whether each object is adopted as a member of Upper Sco in this work.
A similar catalog for young stars within the boundary of Ophiuchus is
presented by \citet{esp20}.
In the remainder of this section, we provide details concerning the
data and membership classifications in Table~\ref{tab:mem}.

\subsection{Membership Classifications}
\label{sec:memclass}

For each star in Table~\ref{tab:mem} that has proper motion and parallax
measurements from {\it Gaia} DR2, we have checked whether it
overlaps at 1~$\sigma$ with the membership criteria in Section~\ref{sec:app}
for Upper Sco, Ophiuchus, or the remainder of Sco-Cen (V1062~Sco, UCL, LCC).
If a parallax measurement is inconsistent at 1--2~$\sigma$ but has
an error of $>$10\%, we ignore that measurement for this analysis.
In addition, if either the parallax or proper motion differs by
$>1$~$\sigma$ from the criteria and the value of RUWE is greater than 1.6,
that measurement is ignored. The same approach was taken in
Section~\ref{sec:app} when identifying candidate members.
If the criteria for a given population are satisfied, a flag for that
population is assigned to the star in Table~\ref{tab:mem}: ``O" for 
Ophiuchus, ``U" for Upper Sco, and ``S" for the remainder of Sco-Cen.
The flag is upper case if the criteria for both proper motion and parallax
are satisfied and it is lower case if the criterion for one parameter
is satisfied while the other parameter has been ignored.
Since the criteria for the three populations
overlap, some stars are consistent with multiple populations.
We also assign a flag of ``n" for stars that have {\it Gaia} data that are
not consistent with any of the three populations.

We have adopted stars as members of Upper Sco if they have
1) flag = U, 2) flag = u and flag $\neq$ S, or 3) flag is absent
(i.e., the necessary {\it Gaia} data are not available).
It is likely that a small minority of the stars in the third category
(mostly late-type objects) are members of UCL/LCC
given that kinematic members of UCL/LCC extend across Upper Sco
on the sky (Figure~\ref{fig:mapsc}).
In Table~\ref{tab:mem}, we indicate whether each star is adopted as a member
of Upper Sco based on these criteria. 
We also have assigned membership in Upper Sco to a small number of stars
that have flag $\neq$ U/u, consisting of six stars for which
parallaxes or proper motions changed significantly between DR1 and DR2
of {\it Gaia}, three objects that are only slightly beyond one of the
kinematic thresholds for membership in Upper Sco, and three stars
that are within a few arcseconds (i.e., likely companions) of
kinematic members of Upper Sco. We have adopted 1761 of the 2020 stars
in Table~\ref{tab:mem} as members of Upper Sco.
We discuss the membership of a few individual stars in the Appendix.

\subsection{Spectral Types and Extinctions}
\label{sec:spt}

Two objects in Table~\ref{tab:mem} lack spectral classifications:
{\it Gaia} 6242434969863194880 has strong IR excess emission and a featureless
near-IR spectrum except for emission lines and $\nu$~Sco~B has not been
observed with spectroscopy. In addition, we do not list adopted
spectral types for Antares, UGCS J161152.78$-$193847.3,
and WISEA J161935.70$-$195043.0. The former is a red supergiant and the latter
two sources have uncertain classifications \citep[][Appendix]{cod17,esp18}.
We have estimated extinctions for stars in Table~\ref{tab:mem} that have 
adopted types using (in order of preference) near-IR spectra, $J-H$,
$J-K_s$ $H-K_s$, and $G_{\rm BP}-G_{\rm RP}$ relative to the typical
intrinsic spectra and colors of young stars \citep[][Appendix]{luh17}.
For companions that lack these data, we adopt the extinction estimates of
their primary stars. The reddenings are converted to extinction in $K_s$ using
relations from \citet{ind05} and \citet{sch16} and 
$E(G_{\rm BP}-G_{\rm RP})/E(J-H)\approx3$, where the latter is 
based on our sample of young stars toward Upper Sco.
Although it has a spectral classification, we have not estimated an
extinction for {\it Gaia} 6243210920137332096 due to the presence of
strong near-IR excess emission.

\subsection{Radial Velocities and $UVW$ Velocities}

In Table~\ref{tab:mem}, we have included previous measurements of radial
velocities that have errors less than 4~km~s$^{-1}$, which are available
for 293 stars. 
We omitted velocities with errors larger than that threshold because they
would have little value in constraining the internal kinematics
of Upper Sco given its velocity dispersion \citep[1--2~km~s$^{-1}$,][]{wri18}.
We adopt errors of 0.4~km~s$^{-1}$ for velocities from
\citet{tor06} for which errors were not reported, which is near
the typical precision estimated in that study.

Among the 293 stars with radial velocities, 270 have {\it Gaia} measurements
of proper motions and parallaxes with errors of $\leq$10\%.
Four of the remaining 23 stars lack counterparts in {\it Gaia} DR2,
all of which are bright enough for detection by {\it Gaia} (e.g., Antares).
The other 19 stars have entries in DR2 and are sufficiently bright 
for precise parallaxes, but most of them have large values of RUWE,
indicating poor astrometric fits.
For 267 of the 270 stars with radial velocities and small relative errors in
proper motion and parallax, we have used the radial velocities,
proper motions, and parallactic distances \citep{bai18}
to compute $UVW$ space velocities \citep{joh87},
which are provided in Table~\ref{tab:mem}.
The errors for the space velocities were estimated
through a Monte Carlo approach in which the realizations were
generated with the python package {\tt pyia} \citep{pri18} using
the radial velocity errors and 
the covariance matrices of errors and correlation coefficients for
the {\it Gaia} astrometry.
Three of the 270 stars (USco~CTIO~55, HD~148184, {\it Gaia}
6242801833086694528) were excluded from the calculations of $UVW$
because their parallaxes are discrepant relative to members of Upper Sco
and may be unreliable based on some combination of a large value of RUWE,
an unusually large parallax error for a given magnitude, or a large change
in parallax measurement between DR1 and DR2 of {\it Gaia}.

Among the 267 stars with $UVW$ estimates, 209 are adopted as members
of Upper Sco in Table~\ref{tab:mem}. 
The median of the space velocities for
those members is $U, V, W = -5.1, -16.0, -7.2$~km~s$^{-1}$, which is
similar to the value derived by \citet{wri18} for a smaller sample of members
that had parallax and proper motion measurements from {\it Gaia} DR1.
The standard deviations of the velocity components are
$\sigma_{UVW}=4.1, 1.4, 2.1$~km~s$^{-1}$. The latter two values are
similar to the intrinsic velocity dispersions in $V$ and $W$ estimated
by \citet{wri18}. Our standard deviation in $U$ is larger than 
the dispersion $U$ from that study because the radial velocity
errors tend to be larger than the proper motion errors for this sample of
stars, and the former contribute primarily to the errors in $U$.

\section{Properties of the Stellar Population}
\label{sec:pop}

\subsection{Initial Mass Function}
\label{sec:imf}

The IMF for Upper Sco has been previously estimated using samples
of a few hundred members \citep{sle08,lod11a,lod13c}. 
The current census of members is much larger and more complete than those
earlier samples, enabling a more accurate characterization of the IMF.
For our analysis, we consider the adopted members from
Table~\ref{tab:mem} that are within the central concentration of stars 
(Figure~\ref{fig:mapsc}), which is where the completeness is greatest.
We define the vertices of that field to be
$(l,b)=(356.8\arcdeg,25.2\arcdeg)$, $(347.4\arcdeg,24.5\arcdeg)$,
and $(356.8\arcdeg,25.2\arcdeg)$.
We exclude Antares and members with uncertain spectral classifications
(Section~\ref{sec:spt}) since we wish to use the distribution of spectral
types as a proxy for the IMF, as done in our previous work
\citep[e.g.,][]{luh16,esp19}.
The resulting sample contains 1422 adopted members.
The field in question also encompasses 204 candidate
members from Table~\ref{tab:cand}. Most of those candidates have not been
observed with spectroscopy, so we have estimated their spectral types
using their $K_s$ photometry and the median relation
between that band and spectral type for the known member 
(i.e., the median of the sequence of members in $K_s$ versus spectral type).
For the small number of candidates that lack IR photometry
(close companions resolved by {\it Gaia}), we have estimated spectral types
with the median relation between $G$ and spectral type.
Most of the brighter candidates with photometry indicative of $\lesssim$M6 have
parallax and proper motion measurements from {\it Gaia}, and thus are likely
to be members.

\begin{figure}
\epsscale{1.2}
\plotone{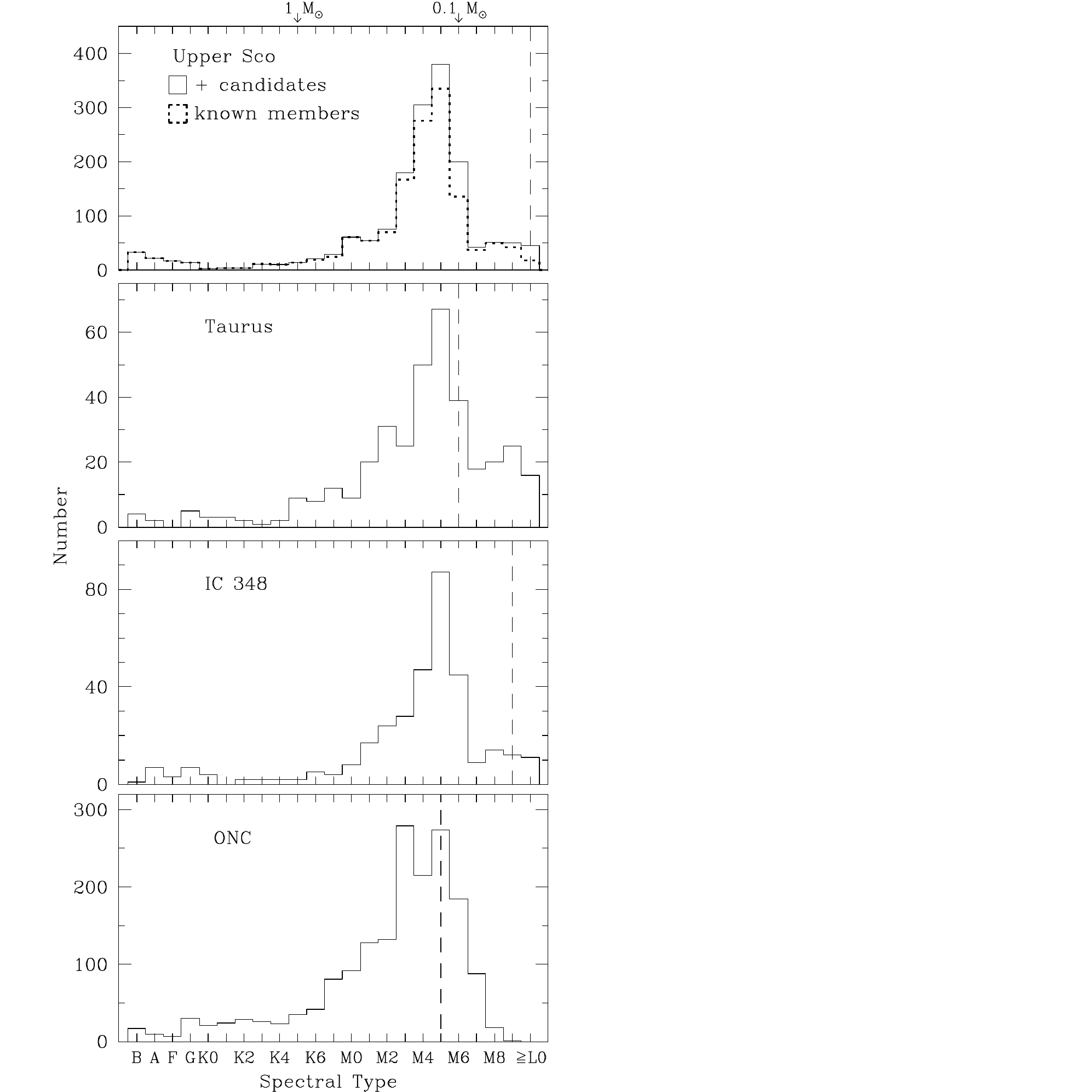}
\caption{
Distributions of spectral types for the central concentration in Upper Sco 
(Figure~\ref{fig:mapsc}),
Taurus \citep[$A_J<1$~mag,][]{esp19},
IC~348 \citep[$A_J<1.5$~mag,][]{luh16}, and the ONC \citep{dar12,hil13}.
We also show the distribution for Upper Sco after including
candidate members that lack confirmation of youth (Table~\ref{tab:cand}).
The dashed lines indicate the completeness limits of these samples
and the arrows mark the spectral types that correspond to masses of 0.1
and 1~$M_\odot$ for ages of a few Myr according to evolutionary models
\citep[e.g.,][]{bar98}.
}
\label{fig:imf}
\end{figure}

In Figure~\ref{fig:imf}, we present the distribution of spectral types
for the adopted members in the central concentration of Upper Sco
and the distribution that combines the adopted and candidate members.
The addition of the candidates has little effect on the shape of the
distribution at $<$M9.
Regardless of whether the candidates are included, the distribution
peaks at M5, which corresponds to masses of $\sim0.15$~$M_\odot$ for the
age of Upper Sco according to evolutionary models \citep{bar98,bar15}.
The shape of the mass function at $\gtrsim$L0 ($\lesssim0.01$~$M_\odot$)
does depend significantly on the fraction of candidates that are members.
Spectroscopy of those candidates will be necessary to better constrain
the IMF in that mass range.

For comparison to Upper Sco, we have included in Figure~\ref{fig:imf}  
the spectral type distributions for extinction-limited samples of
stars in Taurus \citep{esp19} and IC~348 \citep{luh16} and for
a sample of members of the Orion Nebula Cluster (ONC) that was selected
by \citet{luh18tau} with data from \citet{dar12} and \citet{hil13}.
The four regions have similar distributions from $\gtrsim$K4
($\lesssim1$~$M_\odot$) down to their completeness limits, most notably in
terms of the shared peak at M5.
At earlier types, the distributions in Taurus and IC~348 have large
statistical errors because of the small numbers of stars,
but the distributions in Upper Sco and the ONC are sufficiently
populated that they exhibit a significant difference.
The latter is likely a reflection of the evolution of the spectral types
of the high-mass stars between the ages of the ONC and Upper Sco 
\citep[$\sim$2--10~Myr,][]{hil97,reg11,pec12,pec16}.

\subsection{Disk Fraction}
\label{sec:disks}

Mid-IR photometry from {\it WISE} and the {\it Spitzer Space Telescope}
\citep{faz04,rie04,wer04}\footnote{{\it WISE} observed in bands centered at
3.5, 4.5, 12, and 22~$\mu$m, which are denoted as $W1$, $W2$, $W3$, and $W4$,
respectively. {\it Spitzer} obtained images primarily
in bands at 3.6, 4.5, 5.8, 8.0 and 24~\micron, which are denoted
as [3.6], [4.5], [5.8], [8.0], and [24], respectively.}
has been previously used to search for circumstellar disks in Upper Sco
and estimate the fraction of members that harbor disks
\citep{car06,car09,chen11,luh12,ria12,daw13,riz12,riz15,pec16,esp18}.
As with the IMF, we can measure the disk fraction in Upper Sco with
a catalog of members that is larger and more complete than the samples
that were available in previous disk studies.

\citet{luh12} and \citet{esp18} used photometry in three bands from
{\it WISE} ($W2$, $W3$, $W4$) and three bands from {\it Spitzer}
([4.5], [8.0], [24]) to identify excess emission from disks among members of
Upper Sco and to classify the evolutionary stages of the detected disks.
The available {\it Spitzer} images encompass only a subset of the members
of Upper Sco, but they offer better sensitivity than the all-sky data
from {\it WISE}. We have updated the classifications from \citet{luh12} and
\citet{esp18} using our latest extinction estimates and adopted spectral
types (Table~\ref{tab:mem}) and our new estimates for the typical intrinsic
colors of young stellar photospheres (Appendix).
We also have checked for mid-IR excesses among the new stars that now appear
in our membership catalog.
For our adopted members, we present in Table~\ref{tab:disk} a compilation
of photometry from {\it WISE} and {\it Spitzer}, indicators of
excess emission in the six considered bands, and disk classifications.
Data are unavailable in any of those bands for 49 objects, which
are unresolved from brighter stars or below the detection limits.

For their sample of Upper Sco members with mid-IR data,
\citet{luh12} calculated the fraction with excess emission as a function
of spectral type in [4.5]/$W2$, [8.0], $W3$, and [24]/$W4$.
The excess classifications in [4.5]/$W2$ and [24]/$W4$ were combined since the
filters in each pair have similar effective wavelengths (4.5/4.6 and
23.7/22~\micron). We have repeated that analysis 
using our compilation of mid-IR data in Table~\ref{tab:disk}.
Since some of the stars in our catalog for Upper Sco were originally identified
as candidate members via signatures of disks \citep{esp18}, the
current census could be biased in favor of disk-bearing members,
which would mean that the excess fractions computed from that sample are
not representative of the stellar population. To minimize a bias of that
kind, we have included in our excess fractions the 313 candidate members
from Table~\ref{tab:cand} that have parallaxes and proper
motions from {\it Gaia} and photometry from {\it WISE} or {\it Spitzer},
which encompasses most of the candidates that are likely to be earlier
than $\sim$M7.
That sample should have little contamination from non-members based on our
spectroscopy of {\it Gaia}-selected candidates (Section~\ref{sec:spec}).
There remain 65 {\it Gaia} candidates that 
were unresolved from brighter stars in the mid-IR images.
Given that \citet{esp18} already obtained spectra of most candidates for
disk-bearing members from {\it WISE}, only a small number (13)
of the 313 remaining candidates with {\it Gaia} and {\it WISE} data exhibit
possible excess emission.
Most of the excesses are small enough that they are indicative of
debris or evolved transitional disks \citep{esp12}.
We have not attempted to include in the excess fractions
the candidates that are too faint for {\it Gaia} parallaxes because
their membership is less certain.
At the spectral types expected for those faint candidates ($\gtrsim$M9),
only a few members have been identified based on disk signatures
\citep{esp18}, so the sample of known members should not be biased
in terms of disks, and hence their disk fraction should be representative
of the stellar population.

The images in $W2$, [4.5], and [8.0] detected nearly all adopted members of
Upper Sco that were encompassed by them and that were not blended
with brighter stars.
However, some late-type members were too faint for detection in
$W3$, $W4$, and [24], particularly those that lack disks.
As a result, excess fractions in those bands would be biased in favor of
disks at the latest spectral types.
Therefore, as done in \citet{luh12}, we have computed the excess fraction
in each of those bands only for the range of spectral types in which
most members and candidates are detected.
The $W4/[24]$ fraction is based on data from [24] for $\leq$M4 when available
and otherwise uses $W4$ for $\leq$M0. 
There are $\sim$70 stars that are earlier than M4 and have useful
non-detections\footnote{We are referring to non-detections that provide
useful flux limits, namely those that are not due to bright extended emission
or blending with other stars.}
in [24] or are earlier than M0 and have non-detections in $W4$.
All of these non-detections are sufficient to indicate that any excesses
must be quite small ($\lesssim$0.5~mag), corresponding to debris or evolved
transitional disks. We assume that these stars lack excesses for the
calculation of the excess fraction in $W4/[24]$.
The $W3$ excess fraction is computed for stars earlier than M4, among
which a dozen have useful non-detections. As with $W4$ and [24],
those non-detections demonstrate that little excess emission is present,
so we assume that excesses are absent for the $W3$ fractions.

\begin{figure}
\epsscale{1}
\plotone{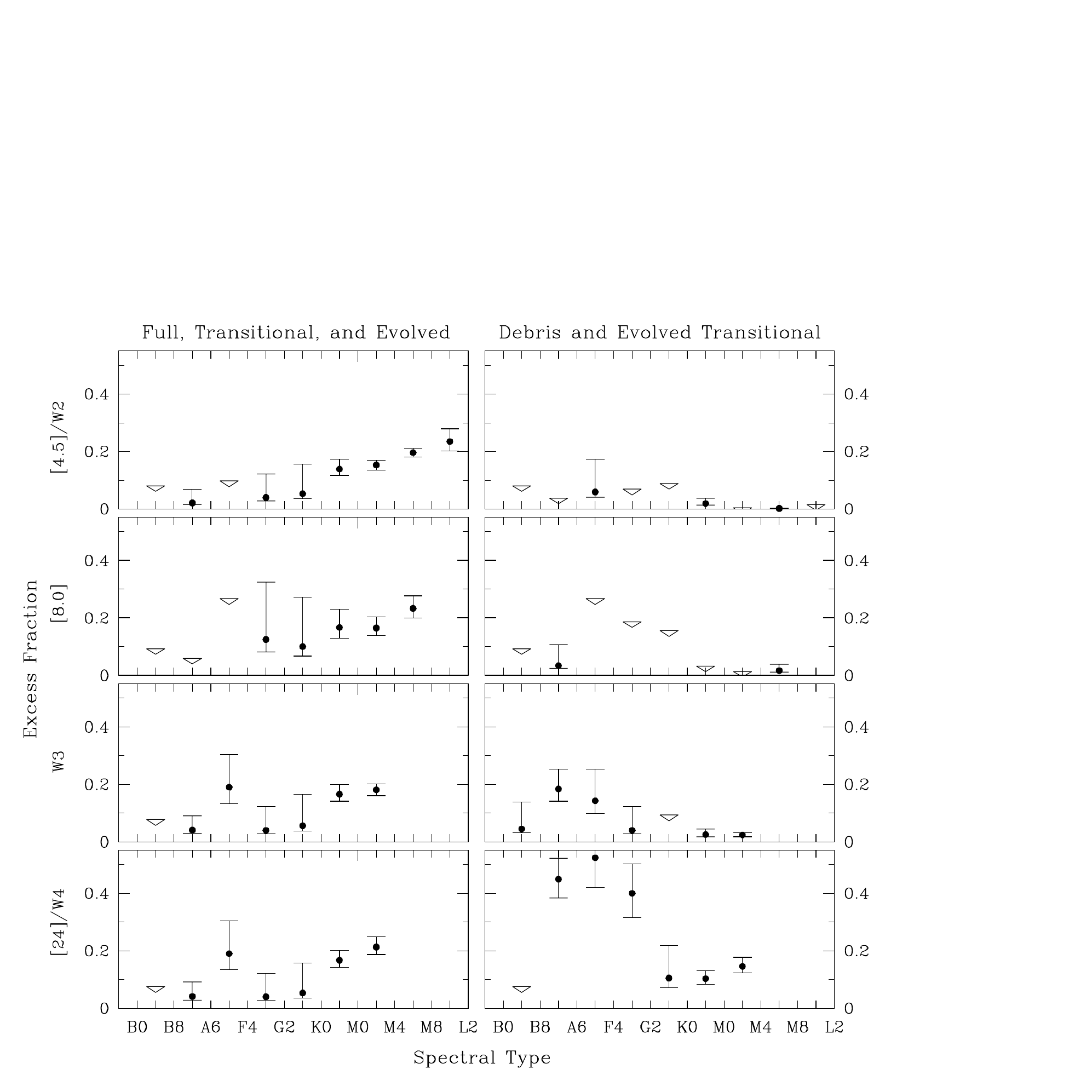}
\caption{
Excess fractions versus spectral type in Upper Sco
for full, transitional, and evolved disks (left) and debris and evolved
transitional disks (right, Table~\ref{tab:fex}).  For each band, data are shown
only down to the coolest spectral type at which most of the known members
are detected. The triangles represent 1~$\sigma$ upper limits.
}
\label{fig:fex}
\end{figure}

We present the excess fractions as a function of spectral type for
[4.5]/$W2$, [8.0], $W3$, and [24]/$W4$ in Table~\ref{tab:fex} and
Figure~\ref{fig:fex}. As done in \citet{luh12}, for each band and spectral
type range we have computed one excess fraction for 
three classes of primordial disks (full, evolved, transitional)
and one fraction for evolved transitional and debris disks.
The latter two classes contain primordial and second-generation dust,
respectively, but they are grouped together because they are indistinguishable
in the available mid-IR data.
The excess fractions are generally similar to those from \citet{car06}
and \citet{luh12} for smaller samples of members.
For the primordial disks, the excess fractions decrease slightly from
longer to shorter wavelengths at a given spectral type, which is a reflection
of clearing of inner disks for the transitional and evolved classes,
and it increases with later spectral types for a given band.
Meanwhile, the excess fraction for debris and evolved transitional disks
is higher for longer wavelengths and earlier types.

\begin{figure}
\epsscale{1}
\plotone{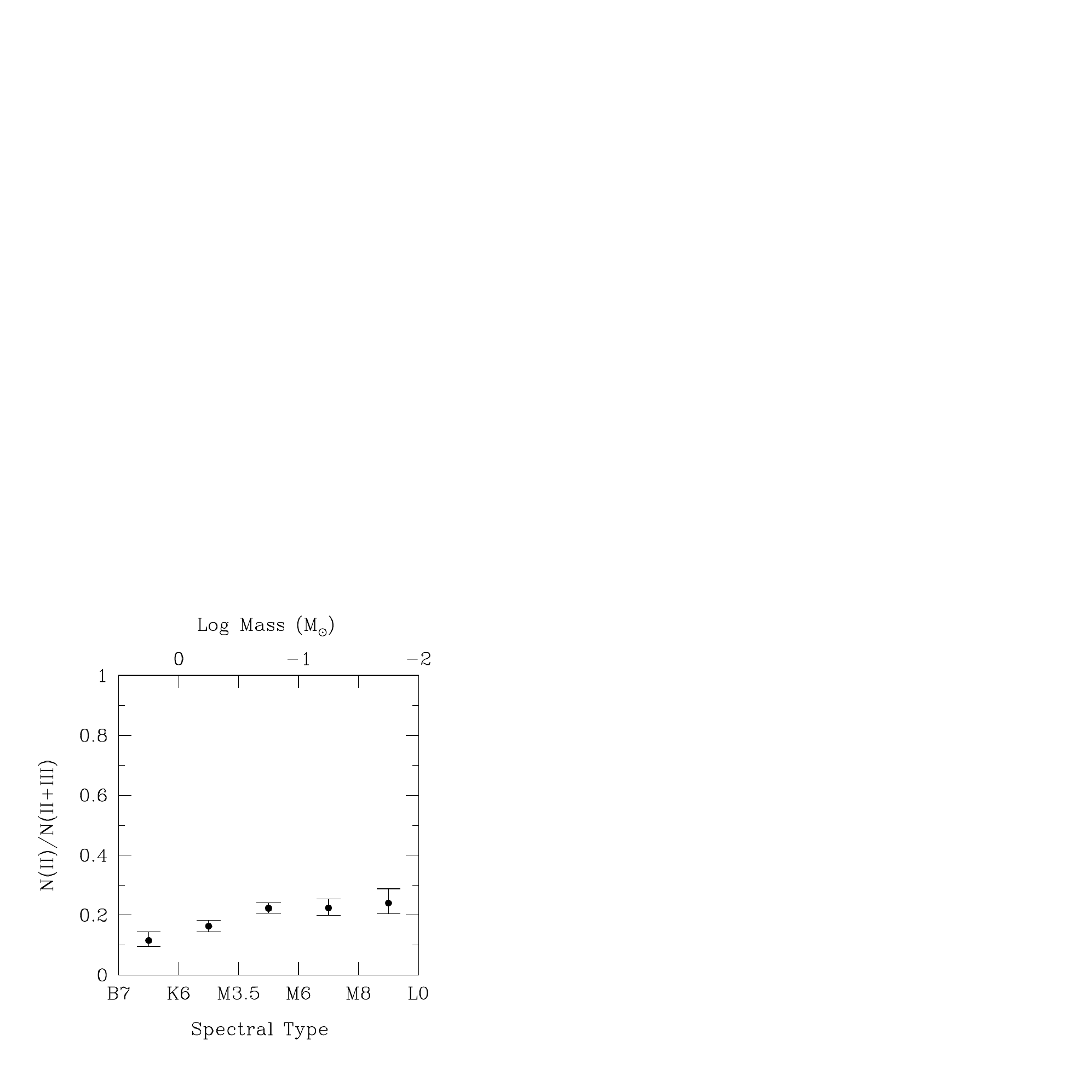}
\caption{
Fraction of Upper Sco members with primordial disks (full, transitional,
evolved) as a function of spectral type (Table~\ref{tab:frac}).
The boundaries of the spectral type bins were chosen to correspond
approximately to logarithmic intervals of mass.
}
\label{fig:frac}
\end{figure}

In addition to the excess fractions for specific bands, we have estimated
the primordial disk fraction as a function of spectral
type based on the fraction of stars that exhibit excess emission in any band 
and that are classified as full, evolved, or transitional.
The results are presented in Table~\ref{tab:frac} and Figure~\ref{fig:frac}.
As done in our previous studies \citep{luh10,luh16,esp19}, we have
derived the fractions for $<$K6, K6--M3.5, M3.75--M5.75, M6--M8, and M8--M9.75,
which approximate logarithmic mass intervals for stars with ages of
$\lesssim10$~Myr \citep{bar98,bar15}. These calculations omit stars that
lack the mid-IR data needed for assessing the presence of excesses. As with 
the excess fractions, the disk fractions increase with later spectral types,
although the variation of disk fraction is more subtle since the highest mass
bin encompasses a wide range of types.

\subsection{Stellar Ages}

Previous studies have estimated the ages of the three subgroups in Sco-Cen
using several diagnostics, 
most notably the main-sequence turnoffs and the empirical isochrones formed
by low-mass members in the Hertzsprung-Russell (H-R) diagram
\citep{pm08,pec12,son12}.
Based on the methods applied to the more massive stars ($\gtrsim1$~$M_\odot$),
\citet{pec16} adopted ages of $\sim$10, 16, and 15~Myr for Upper Sco, UCL, and
LCC, respectively. At lower masses, ages have been derived primarily in Upper 
Sco since it has been searched more extensively for low-mass members
than the other subgroups.  
The ages inferred for the low-mass stars in Upper Sco from the 
H-R diagram have been significantly younger than the ages found at higher masses
\citep[$\sim5$~Myr,][]{deg89,pre02,her15}.
However, evolutionary models that account for the inhibition of convection
by magnetic fields have largely removed this discrepancy by producing
older isochronal ages for the low-mass stars \citep{fei16,mac17,ase19}.
The magnetic models also have reproduced the measured radii of
low-mass eclipsing binaries in Upper Sco for those older ages
\citep{kra15,dav16a,dav19b}.

{\it Gaia} has provided high-precision parallaxes and photometry for 
a large number of low-mass stars in Sco-Cen, which we can use to measure
accurate empirical isochrones in CMDs for the subgroups and estimate
the relative ages of the subgroups and other nearby associations.
To construct these isochrones, we consider a range of colors in which
1) precise parallaxes are available for Sco-Cen ($\pi/\sigma>50$) and 
2) the shapes of theoretical isochrones remain mostly unchanged for
ages of $\sim1$--30~Myr (e.g., stars with different masses experience
similar evolution in their luminosities), which corresponds to
$G_{\rm BP}-G_{\rm RP}=1.4$--2.8~mag ($\sim$0.2--1~$M_\odot$, K5--M4). 
\citet{her15} derived empirical isochrones for young clusters and
associations within a similar range of spectral types.

\begin{figure}
\epsscale{1.2}
\plotone{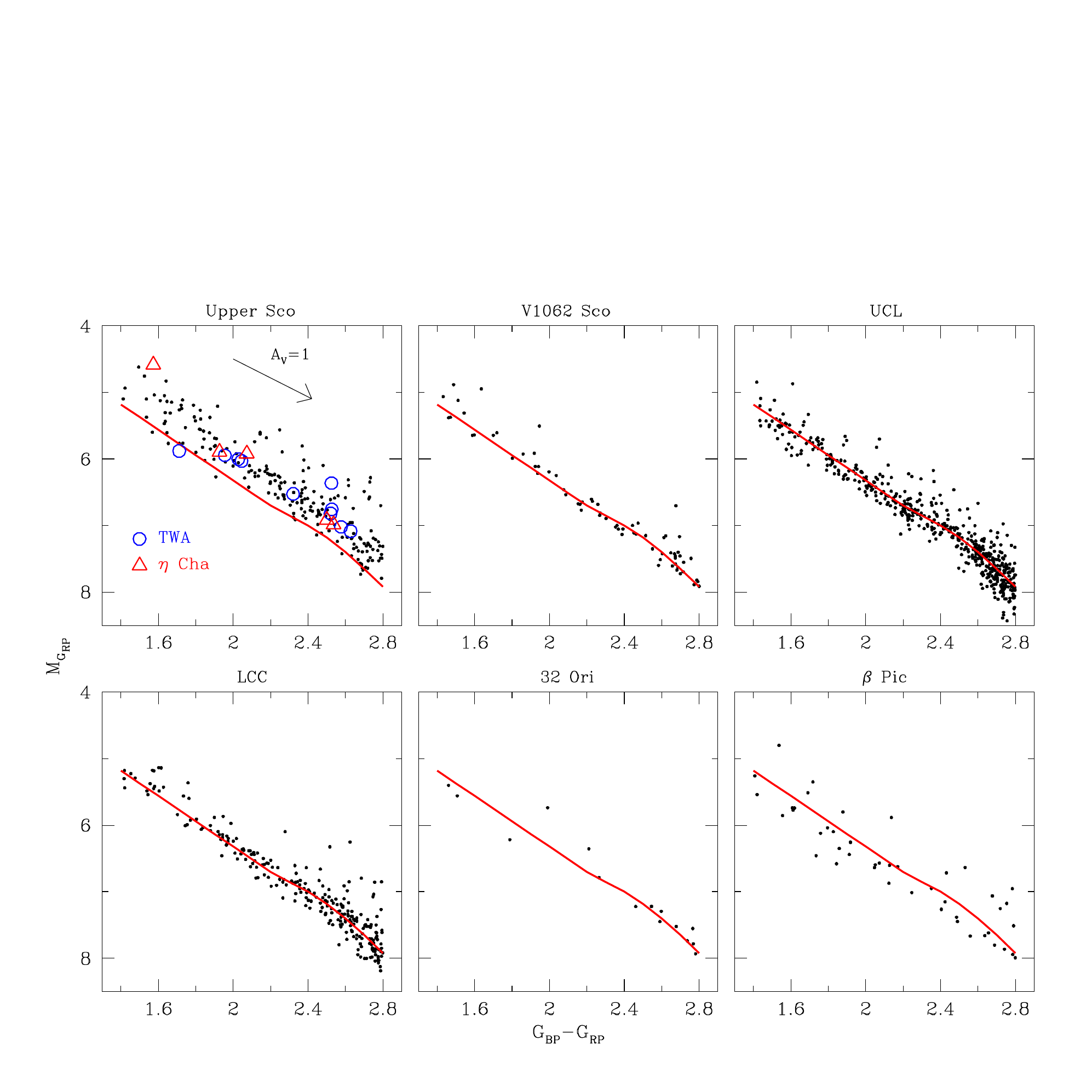}
\caption{
$M_{G_{\rm RP}}$ versus $G_{\rm BP}-G_{\rm RP}$ for low-mass diskless
stars ($\sim0.2$--1~$M_\odot$) in populations within Sco-Cen and in the
32~Ori association \citep{bel17} and the
$\beta$~Pic moving group \citep{bel15,gag18b}.
A fit to the median of the combined sequence for UCL and LCC is shown
with each sample (red solid line).
TWA \citep{gag17} and $\eta$~Cha \citep{luh04,lyo04} are also
plotted with Upper Sco to illustrate the similarity in their ages.
}
\label{fig:br4}

\end{figure}

In Figure~\ref{fig:br4}, we show $M_{G_{\rm RP}}$ versus
$G_{\rm BP}-G_{\rm RP}$ for four samples of stars in Sco-Cen
that are between $G_{\rm BP}-G_{\rm RP}=1.4$--2.8~mag:
1) members of Upper Sco from Table~\ref{tab:mem} that are within the 
central concentration employed for our IMF measurement (Section~\ref{sec:imf})
and have $A_V<1$~mag,
2) stars that are within the 2~$\sigma$ ellipses for the V1062~Sco cluster
from Figure~\ref{fig:pp4} and are within $2\arcdeg$ from V1062~Sco,
3) stars that are within the ellipses for UCL/LCC from Figure~\ref{fig:pp4}
and are within the boundary of UCL from Figure~\ref{fig:mapsc}, and 4) stars
that are within the ellipses for UCL/LCC and are within the boundary
of LCC. Since stars with accretion disks can exhibit excess emission
in short-wavelength bands like $G_{\rm BP}$, we have excluded stars
that have evidence of disks in their mid-IR photometry from {\it Spitzer}
and {\it WISE}.
A typical reddening vector for the intrinsic colors and levels of extinction
of the Upper Sco sample is shown in Figure~\ref{fig:br4}.
Reddening and the sequence in Upper Sco have similar slopes, but they
differ enough for reddening to have a noticeable effect
on the vertical position of the sequence, so we have dereddened the photometry
for Upper Sco using the extinctions from Table~\ref{tab:mem} and have
considered only members with $A_V<1$~mag, as mentioned earlier.
Most known members of UCL and LCC have $A_V<0.5$~mag \citep{pec16}, so we have
not dereddened the data for the other Sco-Cen samples in Figure~\ref{fig:br4}. 

For comparison to the Sco-Cen samples, we have included in Figure~\ref{fig:br4}
members of 32~Ori and $\beta$~Pic because they are 
two of the youngest nearby associations that have ages
estimated with the lithium depletion boundary
\citep[$\sim$21--24~Myr,][]{bin14,bin16,bel15,bel17,shk17}, which is one of 
the most reliable age diagnostics for stellar populations with ages between
20--200~Myr \citep{sod14}. We have adopted the members of 32~Ori from 
\citet{bel17} and the members of $\beta$~Pic from \citet{bel15} and
\citet{gag18d}. We have omitted a few of those stars that are discrepant
in terms of their kinematics and positions in CMDs based on {\it Gaia} data.
Previous studies have found that the TW~Hya and $\eta$~Cha associations
are roughly coeval with Upper Sco \citep[e.g.,][]{her15}, so we have also
plotted their diskless members in the CMD for Upper Sco to enable a direct
comparison. The sequences for TW~Hya and $\eta$~Cha do agree well with
that of Upper Sco, confirming that they have similar ages.

Each of the samples in Figure~\ref{fig:br4} is shown with 
a fit to the median of the combined sequence for UCL and LCC, which
facilitates visual comparison of the sequences.
To compare the sequences quantitatively, we begin 
by computing the offset of each star in $M_{G_{\rm RP}}$ from the fit to
UCL/LCC. We then computed the difference between the median offset for a given
population and the median offset for the $\beta$~Pic sequence.
The error in that difference was characterized using the median absolution
deviation (MAD) for the distribution of differences produced by bootstrapping.
The resulting differences in $M_{G_{\rm RP}}$ relative to $\beta$~Pic
are 0.49$\pm$0.04 (Upper Sco), 0.15$\pm0.06$ (V1062~Sco), 0.09$\pm$0.02
(UCL), 0.14$\pm$0.02 (LCC), and 0.12$\pm$0.06~mag (32~Ori).
Since UCL and LCC appear to be coeval, we also computed the difference
between their combined sequence and $\beta$~Pic, which is 0.10$\pm$0.02~mag.

We have converted the differences in $M_{G_{\rm RP}}$ for
Upper Sco and UCL/LCC to differential ages using the luminosity evolution
predicted by standard non-magnetic models \citep{bar15,cho16,dot16} and 
new versions of the magnetic models from \citet{fei16} provided by
G. Feiden (private communication).
The non-magnetic and magnetic models predict similar changes in luminosity
with age between 10 and 25~Myr.
Relative to $\beta$~Pic, Upper Sco is younger by 0.30$\pm$0.025 dex
(a factor of two) and UCL/LCC is younger by 0.060$\pm$0.024 dex according to
both the non-magnetic and magnetic models.
Based on the lithium depletion boundary in $\beta$~Pic, \citet{bin16}
estimated ages of 21$\pm4$ and 24$\pm$4~Myr with non-magnetic and magnetic
models, respectively. For those ages, our differential measurements imply 
ages of 10.5 and 12~Myr for Upper Sco and ages of 18 and 21~Myr for UCL/LCC.
The latter three values are slightly older than most of the ages reported 
in recent studies of Sco-Cen \citep{pec12,pec16}.
Meanwhile, our analysis indicates that V1062~Sco and 32~Ori have similar ages
as UCL/LCC, although their differential ages have larger errors due to the
smaller samples.
Based on lithium depletion and isochrones, age estimates for 32~Ori and
$\beta$~Pic have agreed to within $\sim1$~Myr and have had errors of
$\pm4$~Myr \citep{mam14,bel17}. However, the sequences in
Figure~\ref{fig:br4} suggest that 32~Ori is younger than $\beta$~Pic by 
$\sim$3~Myr.

\citet{don17} measured parallaxes for 48 K/M stars that were adopted
as members of Upper Sco, 47 of which are also in our catalog of members.
The parallaxes were used to place the stars on the H-R diagram and
ages were estimated from the evolutionary models of \citet{bar15}.
\citet{don17} found that the diskless and disk-bearing stars exhibited
median ages of 3.4 and 6.5~Myr, respectively, which corresponds to a
difference of $\sim$0.45~mag in luminosity.
We have performed a similar age comparison using the new parallaxes and
membership classifications that are now available.
We consider members of Upper Sco from Table~\ref{tab:mem} that have 
$\pi/\sigma>50$, $A_V<1$~mag, RUWE$<$1.6, and spectral types of K5--M4,
which corresponds to a sample of 334 stars, 45 of which have evidence
of disks. Since $G_{\rm BP}$ can be contaminated by accretion-related
emission, we placed the stars in an H-R diagram that uses
spectral type instead of $G_{\rm BP}-G_{\rm RP}$.
As in our earlier analysis, we used extinction-corrected $M_{G_{\rm RP}}$ 
for the vertical axis of the H-R diagram.
We derived a fit to the median of the sequence for our sample
and computed the $M_{G_{\rm RP}}$ offset of each star from the median.
The median offsets of the disk-bearing and diskless samples differ
by only 0.05~mag and both samples have MADs of 0.45~mag. Thus, we find
no significant difference in the ages of diskless and disk-bearing
stars in Upper Sco.

\section{Conclusions}

We have refined the census of stars and brown dwarfs in the Upper Sco
subgroup within the Sco-Cen OB association
using high-precision parallaxes and proper motions from {\it Gaia}
DR2 and optical and IR spectroscopy of candidate members.
Using the resulting catalog of adopted members, we have examined 
the IMF, disk fraction, and median age in Upper Sco.
Our results are summarized as follows:

\begin{enumerate}

\item
We have used a CMD constructed with data from {\it Gaia} DR2 to identify
candidate young stars within the boundary of Sco-Cen from \citet{dez99}.
We fit the kinematics of these stars with a Gaussian mixture model
that contains three components for Sco-Cen, corresponding to Upper
Sco/Ophiuchus/Lupus, the V1062~Sco cluster, and UCL/LCC. We have used the
stars classified as likely members of Sco-Cen within the central concentration
of the Upper Sco association to define kinematic criteria for membership
within that population. The selection of candidate members of Upper Sco
from \citet{luh18} has been updated to include the application of those
kinematic criteria to data from {\it Gaia} DR2.
Our survey for candidates should be complete down to spectral types of
$\sim$L0 ($\sim0.01$~$M_\odot$) for the areas observed by UKIDSS,
which corresponds to most of Upper Sco.

\item
We have analyzed optical and IR spectra of 431 objects toward Upper Sco
to measure their spectral types and check for signatures of youth that
would support their membership, 376 of which are classified as young.
Most of the latter have spectral types of late K through early L
($\sim$0.01--1~$M_\odot$), including 42 that are M8 or later 
($\lesssim0.04$~$M_\odot$).

\item
We have compiled 2020 objects toward Upper Sco that exhibit evidence of youth
in previous studies and this work.
Their membership in Upper Sco has been assessed using available proper
motions and parallaxes, most notably from {\it Gaia} DR2.
Based on that analysis, we have adopted 1761 objects as members of Upper Sco.
There remain 567 candidates from our survey that lack the necessary
spectroscopy for confirming youth, 378 of which have accurate
{\it Gaia} astrometry and hence are probable members.
Thus, Upper Sco likely contains at least $\sim2100$ members.

\item
We have characterized the IMF in Upper Sco in terms of the distribution
spectral types for the central concentration of the association.
The distribution is similar to those in other star-forming regions
like IC~348, Taurus, and the ONC, all of which peak near M5
($\sim$0.15~$M_\odot$).

\item
\citet{luh12} and \citet{esp18} compiled mid-IR photometry from {\it WISE}
and {\it Spitzer} for previous samples of Upper Sco members and used those
data to identify excess emission from disks and to classify the detected disks.
\citet{luh12} also computed the fraction of members exhibiting excesses
as a function of band and spectral type. We have presented an updated
compilation of mid-IR photometry and disk classifications for our 
adopted members. The new excess fractions are qualitatively similar to
those from \citet{luh12}, reproducing a trend originally discovered
by \citet{car06} in which the excess fractions for primordial disks
increase with later spectral types. The fraction of members with disks
ranges from $\lesssim10$\% for $>1$~$M_\odot$ to $\sim22$\% for
0.01--0.3~$M_\odot$.

\item
We have estimated the relative ages of Upper Sco, other populations in
Sco-Cen, and the 32~Ori and $\beta$~Pic associations using their
sequences of low-mass stars in $M_{G_{\rm RP}}$ versus $G_{\rm BP}-G_{\rm RP}$.
According to non-magnetic/magnetic evolutionary models,
the offsets in $M_{G_{\rm RP}}$ among the sequences indicate that
Upper Sco and UCL/LCC are younger than $\beta$~Pic by 0.30 and 0.060~dex,
respectively. \citet{bin16} estimated ages of 21/24$\pm4$~Myr for
the $\beta$~Pic association from the lithium depleting boundary using
non-magnetic/magnetic models, which would imply ages of 10.5/12~Myr for Upper Sco
and 18/21~Myr for UCL and LCC. Although $\beta$~Pic and 32~Ori have been
previously found to have ages that agree to within $\sim1$~Myr (each with
errors of $\pm4$~Myr), their sequences of low-mass stars indicate that 32~Ori
is younger by $\sim3$~Myr.

\end{enumerate}

\acknowledgements
K.L. acknowledges support from NASA grant 80NSSC18K0444 for portions of
this work. We thank Eric Feigelson for advice regarding statistical methods
Gregory Feiden for providing his evolutionary models, and Eric Mamajek
for comments on the manuscript.
The IRTF is operated by the University of Hawaii under contract NNH14CK55B
with NASA. The data at CTIO were obtained through program 2018A-0098
and 2019A-0127 at the National Optical Astronomy Observatory (NOAO).
CTIO and NOAO are operated by the Association of Universities for Research in
Astronomy under a cooperative agreement with the NSF. The Gemini data were 
obtained through programs GN-2018A-Q-115, GN-2018A-Q-218 (NOAO 2018A-0066),
GN-2019A-Q-317 (2019A-0086), GS-2019A-Q-125 (2019A-0053),
and GN-2020A-Q-218 (2020A-0066).
Gemini Observatory is operated by AURA under a cooperative agreement with
the NSF on behalf of the Gemini partnership: the NSF (United States), the NRC
(Canada), CONICYT (Chile), the ARC (Australia),
Minist\'{e}rio da Ci\^{e}ncia, Tecnologia e Inova\c{c}\~{a}o (Brazil) and
Ministerio de Ciencia, Tecnolog\'{i}a e Innovaci\'{o}n Productiva (Argentina).
{\it WISE} is a joint project of the University of California, Los Angeles,
and the JPL/Caltech, funded by NASA.
2MASS is a joint project of the University of Massachusetts and IPAC
at Caltech, funded by NASA and the NSF. This work used data from
the NASA/IPAC Infrared Science Archive, operated by JPL under contract
with NASA, and the SIMBAD database, operated at CDS, Strasbourg, France.
This work has made use of data from the European Space Agency (ESA)
mission {\it Gaia} (\url{https://www.cosmos.esa.int/gaia}), processed by
the {\it Gaia} Data Processing and Analysis Consortium (DPAC,
\url{https://www.cosmos.esa.int/web/gaia/dpac/consortium}). Funding
for the DPAC has been provided by national institutions, in particular
the institutions participating in the {\it Gaia} Multilateral Agreement.
Funding for the Sloan Digital Sky Survey IV has been provided by the Alfred P. Sloan Foundation, the U.S. Department of Energy Office of Science, and the Participating Institutions. SDSS-IV acknowledges
support and resources from the Center for High-Performance Computing at
the University of Utah. The SDSS web site is www.sdss.org.
SDSS-IV is managed by the Astrophysical Research Consortium for the 
Participating Institutions of the SDSS Collaboration including the 
Brazilian Participation Group, the Carnegie Institution for Science, 
Carnegie Mellon University, the Chilean Participation Group, the French Participation Group, Harvard-Smithsonian Center for Astrophysics, 
Instituto de Astrof\'isica de Canarias, The Johns Hopkins University, Kavli Institute for the Physics and Mathematics of the Universe (IPMU) / 
University of Tokyo, the Korean Participation Group, Lawrence Berkeley National Laboratory, Leibniz Institut f\"ur Astrophysik Potsdam (AIP),  
Max-Planck-Institut f\"ur Astronomie (MPIA Heidelberg), 
Max-Planck-Institut f\"ur Astrophysik (MPA Garching), 
Max-Planck-Institut f\"ur Extraterrestrische Physik (MPE), 
National Astronomical Observatories of China, New Mexico State University, 
New York University, University of Notre Dame, 
Observat\'ario Nacional / MCTI, The Ohio State University, 
Pennsylvania State University, Shanghai Astronomical Observatory, 
United Kingdom Participation Group,
Universidad Nacional Aut\'onoma de M\'exico, University of Arizona, 
University of Colorado Boulder, University of Oxford, University of Portsmouth, 
University of Utah, University of Virginia, University of Washington, University of Wisconsin, Vanderbilt University, and Yale University.
The Center for Exoplanets and Habitable Worlds is supported by the
Pennsylvania State University, the Eberly College of Science, and the
Pennsylvania Space Grant Consortium.

\appendix

\section{Comments on Individual Sources}

{\it Gaia} 6045893078810161152 (M0.25) and 6045893074511675008 (M3.75)
comprise a $3\farcs8$ pair.
Their parallaxes and proper motions differ by $>1$~$\sigma$,
but they are roughly similar, which would suggest that they are companions
given their small separation. The first component has a large
enough value of RUWE (2.6) to suggest a poor astrometric fit,
so it may not be surprising that their kinematics do not agree more closely.
However, Li absorption is strong in {\it Gaia} 6045893078810161152  (0.54~\AA)
and absent in {\it Gaia} 6045893074511675008 ($<$0.1~\AA), indicating
that they are not coeval and thus are not companions, or that the latter
has experienced an unusual degree of Li depletion. For the purposes of this
work, we classify {\it Gaia} 6045893074511675008 as a nonmember based
on the old age implied by its lack of Li absorption ($>30$~Myr).

The near-IR spectrum of UGCS J161152.78$-$193847.3 exhibits strong steam
absorption that indicates a spectral type of late M or L and a triangular
$H$-band continuum that indicates a young age \citep{luc01}.
However, the (very red) slope of the spectrum does not agree well with
standard spectra for young M/L objects \citep{luh17}, even when including
reddening as a free parameter. Stars that are observed primarily in scattered 
light (e.g., edge-on disks) can have anomalous spectral slopes at near-IR 
wavelengths \citep{luh07edgeon}.
UGCS J161152.78$-$193847.3 is one of the reddest objects in our catalog of
adopted members of Upper Sco in its IR colors ($J-K_s=2.63$~mag,
$W1-W2=1.07$~mag). It has an uncertain detection in $W3$ and no detection
in $W4$.  These constraints on its IR
colors fall within the ranges of colors observed for young stars with
edge-on disks in Taurus \citep{esp14}. High-resolution imaging and additional 
long-wavelength photometry would be useful for further testing the presence
of an edge-on disk around UGCS J161152.78$-$193847.3.

GSC~06214-00210 is a K7 star that has been previously classified as a
member of Upper Sco \citep{pre98}. 
It harbors a substellar companion at a separation of $2\farcs2$, which
has a mass of $\sim14$~$M_{\rm Jup}$ according to evolutionary models
assuming an age of 5~Myr for Upper Sco \citep{bow11,ire11}.
However, the {\it Gaia} proper motion offset and parallax of GSC~06214-00210
($\Delta\mu_{\alpha,\delta}\sim-6.3,-0.6$~mas~yr$^{-1}$, $\pi=9.188$~mas) 
are inconsistent with the criteria for membership in Upper Sco
in Figure~\ref{fig:pp}, and instead indicate membership in UCL/LCC.
Applying the older age of UCL/LCC to the secondary would result in a
higher estimate of its mass.

EPIC 203710387 is an eclipsing binary that is projected against Upper Sco.
Its components are fainter and smaller in radius than expected for members of
Upper Sco, indicating an older age \citep{dav19b}.
The system's kinematics in Figure~\ref{fig:pp} are consistent with
membership in either Upper Sco or UCL/LCC. 
Given that the kinematic members of UCL/LCC extend across Upper Sco 
(Fig.~\ref{fig:pp}), the data from \citet{dav19b} could be explained by
membership in UCL/LCC.

\clearpage

\section{Extinction Coefficients in {\it Gaia} Bands}
\label{sec:av}

\citet{bab18} derived formulae for extinction coefficients in
$G$, $G_{\rm BP}$, and $G_{\rm RP}$, which were defined as ratios of the
extinctions in those bands relative to the extinction at
5500~\AA\ ($k_\lambda=A_\lambda/A_{5500}$). For broad filters like those
from {\it Gaia} \citep{eva18}, extinction coefficients depend significantly
on the distribution of flux across a filter, which is determined by both the
extinction and intrinsic spectrum of an object.
\citet{bab18} estimated these dependencies by applying a range of extinctions
($A_{5500}=0.01$--5~mag) to spectra over a range of effective temperatures
(3500--10,000~K, $G_{\rm BP}-G_{\rm RP}\sim0$--2.2~mag), adopting the extinction
law from \citet{fit07} for $A_{5500}=3.1 E(B-V)$ and model spectra from 
\citet{cas04}. For each {\it Gaia} band, the resulting extinction coefficients
were fitted by polynomial functions containing $A_{5500}$ and the intrinsic
value of $G_{\rm BP}-G_{\rm RP}$. 

{\it Gaia} is capable of detecting members of nearby star-forming regions
with intrinsic colors and extinctions beyond the limits considered
in the simulation of extinction coefficients by \citet{bab18}.
For instance, {\it Gaia} DR2 provides detections in both $G_{\rm BP}$ and
$G_{\rm RP}$ for members of Upper Sco with spectral types as late as $\sim$M7,
which have intrinsic colors of $G_{\rm BP}-G_{\rm RP}\sim4.3$~mag. 
Therefore, we have derived new formulae for the extinction coefficients
in the {\it Gaia} bands that are applicable to wider ranges of extinctions
and intrinsic colors.
For this analysis, we have adopted the profiles for the {\it Gaia} filters from
\citet{mai18}, the extinction curve from \citet{sch16}
for $x=0$ ($R_V=A_V/E(B-V)\approx3.3$), model spectra for 4000--10,000~K and
log~$g=4$ from \citet{cas04}, and model spectra for 2600--4000~K and log~$g=5$ 
from \citet{bar15}. A lower limit of 2600~K was selected because it corresponds
roughly to our desired limit of $G_{\rm BP}-G_{\rm RP}\sim4.3$~mag.
Like \citet{bab18}, we have fitted the simulated extinction coefficients
with polynomials containing $A_{5500}$ and intrinsic $G_{\rm BP}-G_{\rm RP}$.
Since extinction is more commonly measured in a band rather than at a
single wavelength, we also have fitted the coefficients as a function of
$A_V$ and $G_{\rm BP}-G_{\rm RP}$ where the $V$ filter profile is taken 
from \citet{man15}.
Our adopted formulae for the extinction coefficients are presented in
Table~\ref{tab:red}, which apply to $A_{5500}\leq20$~mag and $A_V\leq20$~mag.
The residuals for these fits are $\lesssim1$\%, $\lesssim2$\%, and 
$\lesssim0.5$\% for $G$, $G_{\rm BP}$, and $G_{\rm RP}$, respectively.

\begin{figure}
\epsscale{1.2}
\plotone{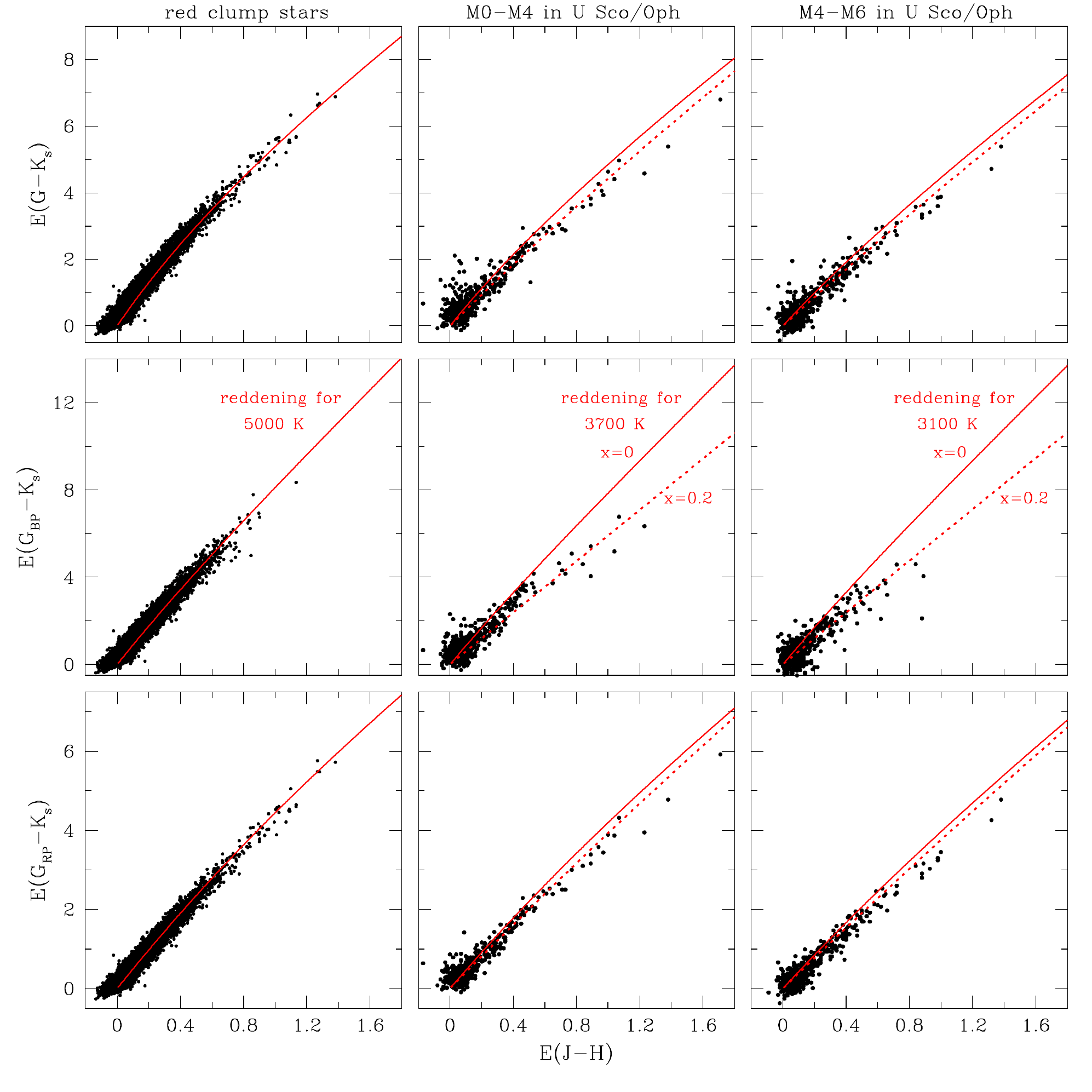}
\caption{
Color excesses in the {\it Gaia} bands relative to $K_s$ versus excesses in
$J-H$ for red clump stars from APOGEE (left) and stars in
Upper Sco and Ophiuchus with spectral types of M0--M4 and M4--M6
that lack excess emission from disks at $<5$~\micron\ (middle and right).
The solid lines represent reddening vectors produced by the
extinction curve from \citet{sch16} for $x=0$ ($R_V\approx3.3$) and model
spectra with intrinsic $G_{\rm BP}-G_{\rm RP}$ colors that are similar to
those of each sample. For the young stars, we also include vectors for
$x=0.2$ ($R_V\approx5$, dotted lines).
}
\label{fig:red}
\end{figure}

The validity of our simulated extinctions can be tested with the
observed colors of reddened stars. We begin by defining a sample of red clump
stars that are subject to varying amounts of extinction in the galactic plane
using spectroscopic data that were obtained through the Apache Point
Observatory Galactic Evolution Experiment \citep[APOGEE,][]{maj17} and that
are available in data release 15 of the Sloan Digital Sky Survey \citep{agu19}.
We have selected APOGEE targets that have 2MASS and {\it Gaia} DR2 counterparts
and that exhibit S/N$>$50, $K_s<14$~mag,
$-0.5\leq[M/H]\leq0$, $4800~K\leq T_{\rm eff}\leq 5000$~K, and
$2.35\leq$log~$g\leq2.55$. The latter three criteria were designed to
encompass a large number of red clump stars while being
narrow enough that the spread in intrinsic colors should be small.
We also omit {\it Gaia} photometry that is fainter than a magnitude of 20
and 2MASS photometry with errors greater than 0.05~mag.
In the left column of Figure~\ref{fig:red}, we plot color excesses in
the {\it Gaia} bands relative to $K_s$ versus excesses in $J-H$ for our
sample of red clump stars. We have computed the excesses by subtracting
intrinsic colors such that the blue end of each sequence of excesses is
aligned with the origin, which consist of $G-K_s=1.95$~mag, 
$G_{\rm BP}-K_s=2.45$~mag,
$G_{\rm RP}-K_s=1.4$~mag, and $J-H=0.52$~mag.
For comparison, we plot our simulated
excesses for $x=0$ ($R_V\approx3.3$) and a model spectrum with
$T_{\rm eff}=5000$~K and log~$g=2.5$. The simulated reddening vector agrees
well with the slope of the observed color excesses.
Such agreement is expected given that the adopted extinction curve from
\citet{sch16} was measured with APOGEE targets in the galactic plane.
To check whether that extinction curve is applicable to Upper Sco and Ophiuchus,
we include in Figure~\ref{fig:red} the color excesses for our compilation
of young stars in those regions that lack excess emission from disks at
$<5$~\micron. Since the reddening vectors
depend on intrinsic stellar colors, we consider only two narrow ranges of
spectral types, M0--M4 and M4--M6, that encompass a large number of stars,
particularly ones at higher extinctions. We plot with those data
our simulated reddening vectors for stars with the intrinsic colors of
young M2 and M5 stars ($G_{\rm BP}-G_{\rm RP}\approx2$ and 3~mag) and for
$x=0$ and $x=2$ ($R_V\approx3.3$ and 5).
For each of the {\it Gaia} bands, the observed color excesses roughly
follow the predicted vector for $x=0$ at low extinctions, but diverge from
it at $E(J-H)\gtrsim0.5$~mag ($A_V\gtrsim5$~mag). All of the latter stars are in
the Ophiuchus clouds. The discrepancy is reduced to varying degrees among
the bands with $x=0.2$, which is qualitatively consistent with the previous
measurements of unusually high values of $R_V$ in Ophiuchus \citep{chi81,cha09}.

We have also examined the dependence of the extinction coefficient
in the 2MASS $K_s$ band ($k_K=A_K/A_V$) on $A_V$ and effective temperature
assuming the extinction curve from \citet{sch16}.
The simulated values of $k_K$ for most temperatures
range from 0.090--0.093 at $A_V\sim0$~mag and from 0.10--0.105 at
$A_V\sim20$~mag. The middle of this sequence of coefficients can be fit
by the following formula: $k_K=0.0916+0.00036A_V+0.0000089A_V^2$.
The residuals from that fit exceed 2\% only for highly reddened cool
stars ($A_V>10$~mag, $<3500$~K).

\section{Intrinsic Colors of Young Stars and Brown Dwarfs}
\label{sec:colors}

By comparing the observed colors of a young star to the intrinsic photospheric
colors expected for its spectral type, one can estimate the extinction
along the line-of-sight to the star or the excess emission from circumstellar
material. In analysis of that kind, the typical colors of main sequence stars
are often adopted for the photospheric colors \citep{ken95}, but some studies
have estimated the intrinsic colors of young stars using members
of nearby associations and star-forming regions 
\citep[][references therein]{luh10,pec13}.
Given the recent availability of photometry from {\it Gaia} and the
continued improvement in membership samples of nearby young populations,
we have derived intrinsic colors of young stars in the {\it Gaia} bands and
updated our estimates of colors in other widely-used bands
from 2MASS, {\it WISE}, and {\it Spitzer}.
We have performed our analysis using known members of 
populations that have ages of $\lesssim20$~Myr
and distances of $\lesssim$150~pc,
consisting of Upper Sco (Section~\ref{sec:classify}),
Taurus \citep[][references therein]{esp19}, and
Chamaeleon~I \citep[][references therein]{esp17}
and the associations containing
TW~Hya \citep{web99,ste99,giz02,zuc04,sch05,loo07,loo10a,loo10b,shk11,kel15,sch12,sch16b,gag14,gag17},
$\eta$~Cha \citep{mam99,luh04,lyo04},
$\beta$~Pic \citep{zuc01,sch10,schl12,shk17},
and 32~Ori \citep{mam07,bel17}.

We have adopted photometry for our sample from the 2MASS Point Source Catalog
($JHK_s$), the AllWISE Source Catalog ($W1$--$W4$), {\it Gaia} DR2 ($G$, 
$G_{\rm BP}$, $G_{\rm RP}$), and the available observations with {\it Spitzer}
\citep[][references therein]{luh10,shv16,esp17,esp18,esp19}.
We also make use of deeper near-IR data that are available from UKIDSS, VISTA
VHS, and dedicated imaging \citep[e.g.,][]{esp17,esp19}. As done in
Section~\ref{sec:app}, the data from UKIDSS and VHS have been adjusted
so that they are calibrated to 2MASS for late-type objects.

Stars in our sample that are within the Local Bubble ($\lesssim100$~pc)
should have very little extinction \citep[$A_V<0.2$~mag,][]{rei11}, so
their intrinsic colors should be reflected in their photometry 
as long as circumstellar disks are absent, which can produce color excesses.
At a given spectral type, the bluest stars in the more distant populations
like Upper Sco and Taurus have similar colors as the stars within the
Local Bubble, indicating that the former also have little extinction
and thus can provide constraints on the intrinsic colors.
When estimating intrinsic colors as a function of spectral type from
these data, we have given preference to stars that have the youngest ages
(in case the colors vary with age) and no evidence of circumstellar disks.
Thus, we have used the data from the $\beta$~Pic and 32~Ori groups
\citep[24~Myr,][]{bel17} only when insufficient data from younger regions
are available for a given color and range of spectral types.

We have not estimated colors involving $G_{\rm BP}$, $W3$, $W4$, and [24]
for some of the latest spectral types because too few objects are
well-detected in those bands. We have not attempted to derive any
intrinsic colors later than L0 given the large uncertainties in spectral types
in that range \citep{luh17}.
Colors relative to some bands were not derived for $\leq$B7 due to the small
number of observations ([3.6], [5.8]) or saturation ($W1$).
When comparing the [4.5] and $W2$ data for our sample, we find no
systematic differences for any of the spectral types in question. 
Therefore, we have combined the data from those two
bands in our analysis. In Table~\ref{tab:intrinsic}, we present our
estimates of the intrinsic colors of young stars for the bands from
2MASS, {\it Gaia}, {\it WISE}, and {\it Spitzer}.
The same values are listed for $K_s-[4.5]$ and $K_s-W2$.
Since [24] and $W4$ are similar bands and the [24] observations are
more sensitive, the values of $K_s-[24]$ are adopted for $K_s-W4$.

\clearpage

\begin{deluxetable}{ll}
\tabletypesize{\scriptsize}
\tablewidth{0pt}
\tablecaption{Candidate Members of Upper Sco\label{tab:cand}}
\tablehead{
\colhead{Column Label} &
\colhead{Description}}
\startdata
2MASS & 2MASS Point Source Catalog source name \\
WISEA & AllWISE Source Catalog source name \\
UGCS & UKIDSS Galactic Clusters Survey source name\tablenotemark{a}\\
Gaia & {\it Gaia} DR2 source name \\
RAdeg & Right Ascension (J2000) \\
DEdeg & Declination (J2000) \\
Ref-Pos & Reference for right ascension and declination\tablenotemark{b} \\
SpType & Spectral type \\
r\_SpType & Spectral type reference\tablenotemark{c} \\
pmRA & Proper motion in right ascension from {\it Gaia} DR2\\
e\_pmRA & Error in pmRA \\
pmDec & Proper motion in declination from {\it Gaia} DR2\\
e\_pmDec & Error in pmDec \\
plx & Parallax from {\it Gaia} DR2\\
e\_plx & Error in plx \\
Gmag & $G$ magnitude from {\it Gaia} DR2\\
e\_Gmag & Error in Gmag \\
GBPmag & $G_{\rm BP}$ magnitude from {\it Gaia} DR2\\
e\_GBPmag & Error in GBPmag \\
GRPmag & $G_{\rm RP}$ magnitude from {\it Gaia} DR2\\
e\_GRPmag & Error in GRPmag \\
RUWE & renormalized unit weight error from \citet{lin18} \\
Jmag & $J$ magnitude \\
e\_Jmag & Error in Jmag \\
Hmag & $H$ magnitude \\
e\_Hmag & Error in Hmag \\
Ksmag & $K_s$ magnitude \\
e\_Ksmag & Error in Ksmag \\
JHKref & $JHK_s$ reference\tablenotemark{d} \\
selection & Selection criteria satisfied by candidate\tablenotemark{e} \\
separation & Angular separation from nearest known young star within $5\arcsec$ \\
compGaia & {\it Gaia} DR2 source name of nearest known young star within $5\arcsec$
\enddata
\tablenotetext{a}{Based on coordinates from DR10 of the
UKIDSS Galactic Clusters Survey for stars with $K_s>10$ from 2MASS.}
\tablenotetext{b}{Sources of the right ascension and declination are 
DR2 of {\it Gaia}, DR10 of the UKIDSS Galactic Clusters Survey, DR6 of
VISTA VHS, and the 2MASS Point Source Catalog.}
\tablenotetext{c}{
(1) \citet{hou88};
(2) \citet{hou82};
(3) this work;
(4) \citet{luh18}.}
\tablenotetext{d}{
2 = 2MASS Point Source Catalog; u = UKIDSS Galactic Clusters Survey DR10;
v = VISTA VHS DR6.}
\tablenotetext{e}{
G/W/i/Y/Z/ip/zp/yp = CMDs in \citet{luh18};
pi = parallax from {\it Gaia} DR2;
gaia/gps/ucac/2m-gaia/2m-ps/ukidss = proper motions in \citet{luh18} and this
work.}
\tablecomments{
The table is available in its entirety in machine-readable form.}
\end{deluxetable}

\begin{deluxetable}{llll}
\tabletypesize{\scriptsize}
\tablewidth{0pt}
\tablecaption{Observing Log\label{tab:log}}
\tablehead{
\colhead{Telescope/Instrument\tablenotemark{a}} &
\colhead{Disperser/Aperture} &
\colhead{Wavelengths/Resolution} &
\colhead{Targets}}
\startdata
Bok Reflector/B\&C & 600~l~mm$^{-1}$/$1\farcs5$ slit & 0.63--0.86~\micron/5~\AA & 24 \\
CTIO 4~m/COSMOS & red VPH/$1\farcs2$ slit & 0.55--0.95~\micron/4~\AA & 177 \\
Gemini North/GNIRS & 31.7~l~mm$^{-1}$/$1\arcsec$ slit & 0.9--2.5~\micron/R=600 & 46 \\
Gemini South/FLAMINGOS-2 & $HK$ grism/0$\farcs$72 slit & 1.10--2.65~\micron/R=450 & 3 \\
Magellan Baade/FIRE & prism/0$\farcs$8 slit & 0.8--2.5~\micron/R=300 & 3 \\
IRTF/SpeX & prism/$0\farcs8$ slit & 0.8--2.5~\micron/R=150 & 179
\enddata
\tablenotetext{a}{
The Gemini Near-Infrared Spectrograph (GNIRS), FLAMINGOS-2, the
Folded-Port Infrared Echellette (FIRE), and SpeX are described by
\citet{eli06}, \citet{eik04}, \citet{sim13}, and \citet{ray03}, respectively.
The Cerro Tololo Ohio State Multi-Object Spectrograph (COSMOS) is
based on an instrument described by \citet{mar11}.}
\end{deluxetable}

\begin{deluxetable}{llllll}
\tabletypesize{\scriptsize}
\tablewidth{0pt}
\tablecaption{Spectroscopic Data for Candidate Members of Upper Sco\label{tab:spec}}
\tablehead{
\colhead{Source Name\tablenotemark{a}} &
\colhead{Spectral Type\tablenotemark{b}} &
\colhead{$W_{\lambda}$(Li)\tablenotemark{c}} &
\colhead{Instrument} &
\colhead{Date} &
\colhead{Young?}\\
\colhead{} &
\colhead{} &
\colhead{(\AA)} &
\colhead{} &
\colhead{} &
\colhead{}}
\startdata
Gaia 6261607772589226496 & M1.5 & 0.50 & COSMOS & 2018 May 31 & Y \\
Gaia 6234768006561947520 & K5 & 0.35 & COSMOS & 2018 May 31 & Y \\
Gaia 6234796284628451968 & M8 & \nodata & GNIRS & 2018 Mar 5 & N? \\
Gaia 6234811166691211008 & M5 & \nodata & SpeX & 2019 Apr 25 & Y \\
UGCS J154519.90$-$261653.3 & M8--L0 & \nodata & GNIRS & 2018 Mar 14 & Y 
\enddata
\tablenotetext{a}{Source names are from DR2 of {\it Gaia},
DR10 of the UKIDSS Galactic Clusters Survey, and the 2MASS Point Source
Catalog.}
\tablenotetext{b}{Uncertainties are 0.25 and 0.5~subclass for optical and
IR spectral types, respectively, unless indicated otherwise.}
\tablenotetext{c}{Typical uncertainties are 0.05~\AA.}
\tablecomments{
The table is available in its entirety in machine-readable form.}
\end{deluxetable}

\clearpage

\LongTables

\begin{deluxetable}{ll}
\tabletypesize{\scriptsize}
\tablewidth{0pt}
\tablecaption{Young Stars Toward Upper Sco\label{tab:mem}}
\tablehead{
\colhead{Column Label} &
\colhead{Description}}
\startdata
2MASS & 2MASS Point Source Catalog source name \\
WISEA & AllWISE Source Catalog source name\tablenotemark{a} \\
UGCS & UKIDSS Galactic Clusters Survey source name\tablenotemark{b}\\
Gaia & {\it Gaia} DR2 source name \\
Name & Other source name \\
RAdeg & Right Ascension (J2000) \\
DEdeg & Declination (J2000) \\
Ref-Pos & Reference for right ascension and declination\tablenotemark{c} \\
SpType & Spectral type \\
r\_SpType & Spectral type reference\tablenotemark{d} \\
Adopt & Adopted spectral type \\
Ak & Extinction in $K_s$ \\
f\_Ak & Method of extinction estimation\tablenotemark{e}\\
pmRA & Proper motion in right ascension from {\it Gaia} DR2\\
e\_pmRA & Error in pmRA \\
pmDec & Proper motion in declination from {\it Gaia} DR2\\
e\_pmDec & Error in pmDec \\
plx & Parallax from {\it Gaia} DR2\\
e\_plx & Error in plx \\
RVel & Radial velocity \\
e\_RVel & Error in RVel \\
r\_RVel & Radial velocity reference\tablenotemark{f} \\
U & $U$ component of space velocity \\
e\_U & Error in U \\
V & $V$ component of space velocity \\
e\_V & Error in V \\
W & $W$ component of space velocity \\
e\_W & Error in W \\
Gmag & $G$ magnitude from {\it Gaia} DR2\\
e\_Gmag & Error in Gmag \\
GBPmag & $G_{\rm BP}$ magnitude from {\it Gaia} DR2\\
e\_GBPmag & Error in GBPmag \\
GRPmag & $G_{\rm RP}$ magnitude from {\it Gaia} DR2\\
e\_GRPmag & Error in GRPmag \\
RUWE & renormalized unit weight error from \citet{lin18} \\
Jmag & $J$ magnitude \\
e\_Jmag & Error in Jmag \\
Hmag & $H$ magnitude \\
e\_Hmag & Error in Hmag \\
Ksmag & $K_s$ magnitude \\
e\_Ksmag & Error in Ksmag \\
JHKref & $JHK_s$ reference\tablenotemark{g} \\
Pops & {\it Gaia} parallax and proper motion consistent with these populations\tablenotemark{h} \\
USco & Adopted member of Upper Sco?
\enddata
\tablenotetext{a}{The following names are from the WISE All-Sky Source Catalog:
J160027.15$-$223850.5, J160414.16$-$212915.5, J161320.78$-$175752.3,
J161317.38$-$292220.0, J161837.22$-$240522.8, J162210.14$-$240905.4,
J162620.15$-$223312.8.}
\tablenotetext{b}{Based on coordinates from DR10 of the
UKIDSS Galactic Clusters Survey for stars with $K_s>10$ from 2MASS.}
\tablenotetext{c}{Sources of the right ascension and declination are DR2 of
{\it Gaia}, DR10 of the UKIDSS Galactic Clusters Survey, DR6 of
VISTA VHS, the 2MASS Point Source Catalog, and high-resolution imaging 
\citep{ire11,lac15,bry16}.}
\tablenotetext{d}{
(1) \citet{hou88};
(2) \citet{luh18};
(3) \citet{kun99};
(4) \citet{pre98};
(5) \citet{pec16};
(6) \citet{hil69};
(7) this work;
(8) \citet{tor06};
(9) \citet{esp18};
(10) \citet{riz15};
(11) \citet{daw14};
(12) \citet{lod06};
(13) \citet{lod08};
(14) \citet{bon14};
(15) \citet{rui87};
(16) \citet{bes17};
(17) \citet{vie03};
(18) \citet{pen16};
(19) \citet{hou82};
(20) \citet{pec12};
(21) \citet{cor84};
(22) \citet{shk09};
(23) \citet{cru03};
(24) measured in this work with the most recently published spectrum;
(25) \citet{all13b};
(26) \citet{ard00};
(27) \citet{mar04};
(28) \citet{wal94};
(29) \citet{mar10};
(30) \citet{kra09};
(31) \citet{sle08};
(32) \citet{chi20};
(33) \citet{pre02};
(34) \citet{mor01};
(35) \citet{rei08};
(36) \citet{kir10};
(37) \citet{all13};
(38) \citet{fah16};
(39) \citet{giz02};
(40) \citet{her14};
(41) \citet{sle06};
(42) \citet{ria06};
(43) \citet{lod18};
(44) \citet{kra15};
(45) \citet{mac12};
(46) \citet{cod17};
(47) \citet{laf11};
(48) \citet{lac15};
(49) \citet{pre01};
(50) \citet{dav19b};
(51) \citet{ans16};
(52) \citet{pra02};
(53) \citet{bej08};
(54) \citet{her09};
(55) \citet{kra07};
(56) \citet{lod11a};
(57) \citet{ase19};
(58) \citet{bil11};
(59) \citet{laf08};
(60) \citet{luh17};
(61) \citet{man16};
(62) \citet{dav16b};
(63) \citet{sta17};
(64) \citet{sta18};
(65) \citet{coh79};
(66) \citet{pra03};
(67) \citet{eis05};
(68) \citet{gar67};
(69) \citet{mur69};
(70) \citet{mar98a};
(71) \citet{pra07};
(72) \citet{cow69};
(73) \citet{mcc10};
(74) \citet{lod15};
(75) \citet{dav16a};
(76) \citet{luh05usco};
(77) \citet{mar98b};
(78) \citet{esp20};
(79) \citet{gra06};
(80) \citet{lut77};
(81) \citet{cie10};
(82) \citet{bow11};
(83) \citet{bow14};
(84) \citet{rom12};
(85) \citet{jay06};
(86) \citet{clo07};
(87) \citet{luh07};
(88) \citet{mer10};
(89) \citet{dav17};
(90) \citet{tuc15};
(91) \citet{bra97};
(92) \citet{bou92};
(93) \citet{bow17}.}
\tablenotetext{e}{
Extinction estimated from a near-IR spectrum or the indicated color
assuming the intrinsic spectrum from \citet{luh17} or the intrinsic color
from the Appendix for the spectral type in question.}
\tablenotetext{f}{
(1) \citet{dah12};
(2) {\it Gaia} DR2;
(3) \citet{jil06};
(4) \citet{gon06};
(5) \citet{tor06};
(6) \citet{chen11}.
(7) \citet{kur06};
(8) \citet{kun17};
(9) \citet{dur18};
(10) \citet{and83};
(11) \citet{muz03};
(12) \citet{dav16a};
(13) \citet{whi07};
(14) \citet{mac12};
(15) \citet{pra02};
(16) \citet{ric10};
(17) \citet{alo15};
(18) \citet{dav16b};
(19) \citet{eis05};
(20) \citet{wan18};
(21) \citet{dav17}.}
\tablenotetext{g}{
2 = 2MASS Point Source Catalog; u = UKIDSS Galactic Clusters Survey DR10;
v = VISTA VHS DR6; van = \citet{van96}; duc = \citet{duc02};
luh = \citet{luh07}; bow = \citet{bow11}; kra = \citet{kra14};
lac = \citet{lac15}.}
\tablenotetext{h}{O/o = Ophiuchus; U/u = Upper Sco; S/s = remainder of Sco-Cen;
n = none of those populations. Upper case letters
indicate that both proper motion and parallax support membership
in that population. Lower case letters indicate that either proper motion
or parallax supports membership while the other parameter is inaccurate,
unreliable, or unavailable.
}
\tablecomments{
The table is available in its entirety in machine-readable form.}
\end{deluxetable}

\newpage

\begin{deluxetable}{ll}
\tabletypesize{\scriptsize}
\tablewidth{0pt}
\tablecaption{Mid-IR Photometry for Adopted Members of Upper Sco\label{tab:disk}}
\tablehead{
\colhead{Column Label} &
\colhead{Description}}
\startdata
2MASS & 2MASS Point Source Catalog source name \\
WISEA & AllWISE Source Catalog source name \\
UGCS & UKIDSS Galactic Clusters Survey source name \\
Gaia & {\it Gaia} DR2 source name \\
Name & Other source name \\
RAdeg & Right Ascension (J2000) \\
DEdeg & Declination (J2000) \\
Ref-Pos & Reference for right ascension and declination \\
3.6mag & {\it Spitzer} [3.6] magnitude \\
e\_3.6mag & Error in 3.6mag \\
f\_3.6mag & Flag on 3.6mag\tablenotemark{a} \\
4.5mag & {\it Spitzer} [4.5] magnitude \\
e\_4.5mag & Error in 4.5mag \\
f\_4.5mag & Flag on 4.5mag\tablenotemark{a} \\
5.8mag & {\it Spitzer} [5.8] magnitude \\
e\_5.8mag & Error in 5.8mag \\
f\_5.8mag & Flag on 5.8mag\tablenotemark{a} \\
8.0mag & {\it Spitzer} [8.0] magnitude \\
e\_8.0mag & Error in 8.0mag \\
f\_8.0mag & Flag on 8.0mag\tablenotemark{a} \\
24mag & {\it Spitzer} [24] magnitude \\
e\_24mag & Error in 24mag \\
f\_24mag & Flag on 24mag\tablenotemark{a} \\
W1mag & {\it WISE} $W1$ magnitude \\
e\_W1mag & Error in W1mag \\
f\_W1mag & Flag on W1mag\tablenotemark{a} \\
W2mag & {\it WISE} $W2$ magnitude \\
e\_W2mag & Error in W2mag \\
f\_W2mag & Flag on W2mag\tablenotemark{a} \\
W3mag & {\it WISE} $W3$ magnitude \\
e\_W3mag & Error in W3mag \\
f\_W3mag & Flag on W3mag\tablenotemark{a} \\
W4mag & {\it WISE} $W4$ magnitude \\
e\_W4mag & Error in W4mag \\
f\_W4mag & Flag on W4mag\tablenotemark{a} \\
Exc4.5 & Excess present in [4.5]? \\
Exc8.0 & Excess present in [8.0]? \\
Exc24 & Excess present in [24]? \\
ExcW2 & Excess present in $W2$? \\
ExcW3 & Excess present in $W3$? \\
ExcW4 & Excess present in $W4$? \\
DiskType & Disk Type
\enddata
\tablecomments{This table is available in its entirety in a machine-readable form.}
\tablenotetext{a}{nodet = non-detection; sat = saturated; out = outside of the
          camera's field of view; bl = photometry may be affected by
          blending with a nearby star; ext = photometry is known or
          suspected to be contaminated by extended emission (no data
          given when extended emission dominates); dif = photomtery
          may be affected by a diffraction spike; bin = includes an
          unresolved binary companion; unres = too close to a brighter
          star to be detected; false = detection from WISE catalog
          appears false or unreliable based on visual inspection;
          off = $W3$ and/or $W4$ detection appears offset from the $W1/W2$
          detection (no data given when offset is due to a known
          source); err = $W2$ magnitudes brighter than $\sim$6 are
          erroneous.}
\end{deluxetable}

\clearpage

\begin{deluxetable}{llllll}
\tablecolumns{6}
\tabletypesize{\scriptsize}
\tablewidth{0pt}
\tablecaption{Excess Fractions in Upper Sco\label{tab:fex}}
\tablehead{
\colhead{Spectral Type} &
\colhead{Mass\tablenotemark{a}} &
\colhead{[4.5]/W2} &
\colhead{[8.0]} &
\colhead{W3} &
\colhead{[24]/W4} \\
\colhead{} &
\colhead{($M_\odot$)} &
\colhead{} &
\colhead{} &
\colhead{} &
\colhead{}
}
\startdata
\cutinhead{Full, Transitional, and Evolved Disks}
B0--B8 & 2.8--18 & 0/21=$<$0.08 & 0/18=$<$0.09 & 0/22=$<$0.08 & 0/23=$<$0.07\\
B8--A6 & 1.8--2.8 & 1/47=$0.02^{+0.05}_{-0.01}$ & 0/29=$<$0.06 & 2/49=$0.04^{+0.05}_{-0.01}$ & 2/49=$0.04^{+0.05}_{-0.01}$\\
A6--F4 & 1.5--1.8 & 0/17=$<$0.10 & 0/5=$<$0.27 & 4/21=$0.19^{+0.11}_{-0.06}$ & 4/21=$0.19^{+0.11}_{-0.06}$\\
F4--G2 & 1.4--1.5 & 1/25=$0.04^{+0.08}_{-0.01}$ & 1/8=$0.12^{+0.20}_{-0.04}$ & 1/25=$0.04^{+0.08}_{-0.01}$ & 1/25=$0.04^{+0.08}_{-0.01}$\\
G2--K0 & 1.3--1.4 & 1/19=$0.05^{+0.10}_{-0.02}$ & 1/10=$0.10^{+0.17}_{-0.03}$ & 1/18=$0.06^{+0.11}_{-0.02}$ & 1/19=$0.05^{+0.10}_{-0.02}$\\
K0--M0 & 0.7--1.3 & 22/158=$0.14^{+0.03}_{-0.02}$ & 9/54=$0.17^{+0.06}_{-0.04}$ & 26/157=$0.17^{+0.03}_{-0.02}$ & 26/156=$0.17\pm0.03$\\
M0--M4 & 0.2--0.7 & 86/563=$0.15\pm0.02$ & 21/127=$0.17^{+0.04}_{-0.03}$ & 97/537=$0.18\pm0.02$ & 38/178=$0.21\pm0.03$\\
M4--M8 & 0.035--0.2 & 203/1036=$0.20\pm0.01$ & 27/116=$0.23^{+0.04}_{-0.03}$ & \nodata & \nodata\\
M8--L2 & 0.01--0.035 & 27/115=$0.23^{+0.04}_{-0.03}$ & \nodata & \nodata & \nodata\\
\cutinhead{Debris and Evolved Transitional Disks}
B0--B8 & 2.8--18 & 0/21=$<$0.08 & 0/18=$<$0.09 & 1/22=$0.05^{+0.09}_{-0.01}$ & 0/23=$<$0.07\\
B8--A6 & 1.8--2.8 & 0/47=$<$0.04 & 1/29=$0.03^{+0.07}_{-0.01}$ & 9/49=$0.18^{+0.07}_{-0.04}$ & 22/49=$0.45\pm0.07$\\
A6--F4 & 1.5--1.8 & 1/17=$0.06^{+0.11}_{-0.02}$ & 0/5=$<$0.27 & 3/21=$0.14^{+0.11}_{-0.04}$ & 11/21=$0.52\pm0.10$\\
F4--G2 & 1.4--1.5 & 0/25=$<$0.07 & 0/8=$<$0.19 & 1/25=$0.04^{+0.08}_{-0.01}$ & 10/25=$0.40^{+0.10}_{-0.08}$\\
G2--K0 & 1.3--1.4 & 0/19=$<$0.09 & 0/10=$<$0.16 & 0/18=$<$0.09 & 2/19=$0.11^{+0.11}_{-0.03}$\\
K0--M0 & 0.7--1.3 & 3/158=$0.02^{+0.02}_{-0.01}$ & 0/54=$<$0.03 & 4/157=$0.03^{+0.02}_{-0.01}$ & 16/156=$0.10^{+0.03}_{-0.02}$\\
M0--M4 & 0.2--0.7 & 1/563=$0.002^{+0.004}_{-0.001}$ & 0/127=$<$0.01 & 13/537=$0.02\pm0.01$ & 26/178=$0.15^{+0.03}_{-0.02}$\\
M4--M8 & 0.035--0.2 & 2/1036=$0.002^{+0.002}_{-0.001}$ & 2/116=$0.02^{+0.02}_{-0.01}$ & \nodata & \nodata\\
M8--L2 & 0.01--0.035 & 0/115=$<$0.02 & \nodata & \nodata & \nodata
\enddata
\tablenotetext{a}{Masses that correspond to the given range of spectral types
for an age of 10~Myr \citep{bar98,bar15,cho16,dot16}.}
\end{deluxetable}

\clearpage

\begin{deluxetable}{ll}
\tabletypesize{\scriptsize}
\tablewidth{0pt}
\tablecaption{Disk Fraction for Upper Sco
\label{tab:frac}}
\tablehead{
\colhead{Spectral Type} &
\colhead{N(primordial disks)/N(all stars)}}
\startdata
$<$K6 & 20/174=$0.11^{+0.03}_{-0.02}$\\
K6--M3.5 & 84/515=$0.16\pm0.02$\\
M3.75--M5.75 & 215/966=$0.22\pm0.02$\\
M6--M8 & 51/228=$0.22\pm0.03$ \\
$>$M8--M9.75 & 27/107=$0.25^{+0.05}_{-0.04}$
\enddata
\end{deluxetable}

\clearpage

\begin{deluxetable}{lrrrrrrrrr}
\tabletypesize{\scriptsize}
\tablewidth{0pt}
\tablecaption{Parameters for Extinction Coefficients in {\it Gaia} Bands\tablenotemark{a}\label{tab:red}}
\tablehead{
\colhead{} &
\colhead{$c_1$} &
\colhead{$c_2$} &
\colhead{$c_3$} &
\colhead{$c_4$} &
\colhead{$c_5$} &
\colhead{$c_6$} &
\colhead{$c_7$} &
\colhead{$c_8$} &
\colhead{$c_9$}}
\startdata
$k_{G,5500}$  & 0.9605 & $-$0.1355 & $-$0.03399 & 0.00532 & 0.00847 & 0.001217 & 0.000464 & $-$0.0000199 & $-$0.0000257\\
$k_{G,V}$ & 0.9536 & $-$0.1145 & $-$0.03211 & 0.00554 & 0.00000 & 0.001242 & 0.001398 & $-$0.0000208 & $-$0.0000235\\
$k_{BP,5500}$ & 1.1522 & $-$0.1150 & $-$0.02237 & 0.00296 & 0.02487 & 0.000730 & $-$0.001754 & $-$0.0000097 & $-$0.0000197\\
$k_{BP,V}$ & 1.1425 & $-$0.0900 & $-$0.01960 & 0.00322 & 0.01663 & 0.000797 & $-$0.001016 & $-$0.0000104 & $-$0.0000142\\
$k_{RP,5500}$ & 0.6389 & $-$0.0189 & $-$0.00782 & 0.00075 & $-$0.00744 & 0.000081 & 0.001150 & 0.0000002 & $-$0.0000001\\
$k_{RP,V}$ & 0.6309 & $-$0.0005 & $-$0.00540 & 0.00060 & $-$0.01363 & 0.000045 & 0.001724 & 0.0000014 & 0.0000073
\enddata
\tablenotetext{a}{$k_{\lambda 1,\lambda 2}=A_{\lambda 1}/A_{\lambda 2}=c_1+c_2(G_{\rm BP}-G_{\rm RP})+c_3A_{\lambda 2}+c_4(G_{\rm BP}-G_{\rm RP})A_{\lambda 2}+c_5(G_{\rm BP}-G_{\rm RP})^2+c_6A_{\lambda 2}^2+c_7(G_{\rm BP}-G_{\rm RP})^3+c_8A_{\lambda 2}^3+c_9(G_{\rm BP}-G_{\rm RP})^2A_{\lambda 2}^2$ where
$G_{\rm BP}-G_{\rm RP}$ is the intrinsic color. These coefficients
are based on the extinction curve from \citet{sch16} for $x=0$ 
($R_V\approx3.3$) and are valid for stars with intrinsic colors of
$G_{\rm BP}-G_{\rm RP}\sim0$--4.4~mag
and $A_V\leq20$~mag and $A_{5500}\leq20$~mag.}
\end{deluxetable}

\clearpage

\begin{deluxetable}{ll}
\tabletypesize{\scriptsize}
\tablewidth{0pt}
\tablecaption{Intrinsic Colors of Young Stars and Brown Dwarfs\label{tab:intrinsic}}
\tablehead{
\colhead{Column Label} &
\colhead{Description}}
\startdata
SpType & Spectral Type \\
BP$-$RP & $G_{\rm BP}-G_{\rm RP}$ for {\it Gaia} DR2 bands \\
G$-$RP & $G-G_{\rm RP}$ for {\it Gaia} DR2 bands \\
RP$-$K & $G_{\rm RP}-K_s$  for {\it Gaia} DR2 and 2MASS bands \\
J$-$H & $J-H$ for 2MASS bands \\
H$-$K & $H-K_s$ for 2MASS bands \\
K$-$3.6 & $K_s-[3.6]$ for 2MASS and {\it Spitzer} bands \\
K$-$4.5 & $K_s-[4.5]$ for 2MASS and {\it Spitzer} bands\tablenotemark{a} \\
K$-$5.8 & $K_s-[5.8]$ for 2MASS and {\it Spitzer} bands \\
K$-$8.0 & $K_s-[8.0]$ for 2MASS and {\it Spitzer} bands \\
K$-$24 & $K_s-[24]$ for 2MASS and {\it Spitzer} bands\tablenotemark{b} \\
K$-$W1 & $K_s-W1$ for 2MASS and {\it WISE} bands \\
K$-$W2 & $K_s-W2$ for 2MASS and {\it WISE} bands\tablenotemark{a}\\
K$-$W3 & $K_s-W3$ for 2MASS and {\it WISE} bands \\
K$-$W4 & $K_s-W4$ for 2MASS and {\it WISE} bands\tablenotemark{b}
\enddata
\tablenotetext{a}{Same values listed for $K_s-[4.5]$ and $K_s-W2$.}
\tablenotetext{b}{Same values listed for $K_s-[24]$ and $K_s-W4$.}
\tablecomments{
The table is available in its entirety in machine-readable form.}
\end{deluxetable}

\clearpage


\end{document}